\documentclass[usenatbib,fleqn,useAMS]{mnras} 
\usepackage{epsfig}
\usepackage{newtxmath}
\usepackage{newtxtext}
\usepackage{fixltx2e} 
\usepackage{amsmath,amssymb}

\newcommand{\Msol}{{\,\rm M}_\odot} \newcommand{\Mpc} {{\,\rm Mpc}}
\newcommand{\kpc} {{\,\rm kpc}} \newcommand{\Gyr} {{\,\rm Gyr}}
\newcommand{\pc} {{\,\rm pc}} 
\newcommand{\kms}{{\,\rm {km\,s^{-1}} }}

\usepackage{hyperref}

\def\Gyr{\,{\rm Gyr}}

\newcommand{\cc}{{\,\rm {cm^{-3}}}}

\newcommand{\vel}{\mbox{$v$}}

\newcommand{\report}[1]{{{#1}}}

\def\cs {\rm{cm}^{2}}       
\def\kms {\rm{km} \, \rm{s}^{-1}}       
\def\msunyr {\rm M_{\odot} \, yr^{-1}}       

\def\eavg {\bar{\epsilon}}   
\def\csavg {\sigma}           

\def\h2    {{\rm{H$_2$}}}
\def\h    {{\rm{H}}}

\def\hi    {{\rm{H\textsc{i}}}}

\def\hei    {{\rm{He\textsc{i}}}}
\def\heii    {{\rm{He\textsc{ii}}}}
\def\heiii    {{\rm{He\textsc{iii}}}}

\def\ramses    {{\sc Ramses}}
\def\ramsesrt  {{\sc Ramses-RT}}

\def\cloudy    {{\sc Cloudy}}

\def\ramses    {{\sc Ramses}}
\def\ramsesrt  {{\sc Ramses-RT}}

\def\ltsima{$\; \buildrel < \over \sim \;$}
\def\simlt{\lower.5ex\hbox{\ltsima}}
\def\gtsima{$\; \buildrel > \over \sim \;$}
\def\simgt{\lower.5ex\hbox{\gtsima}}

\title[Mass-metallicity relation of ultra-faints]{
EDGE: the mass-metallicity relation as a critical test of galaxy formation physics}

\author[Oscar Agertz et al.] 
{Oscar Agertz$^{1}$\thanks{\tt oscar.agertz@astro.lu.se},
Andrew Pontzen$^{2}$, 
Justin I. Read$^{3}$, 
Martin P. Rey$^{2}$, 
Matthew Orkney$^{3}$, 
\newauthor
Joakim Rosdahl$^4$, 
Romain Teyssier$^5$,
Robbert Verbeke$^5$,
Michael Kretschmer$^5$ and
\newauthor
Sarah Nickerson$^5$
\\
$^1$Lund Observatory, Department of Astronomy and Theoretical Physics, Lund University, Box 43, SE-221 00 Lund, Sweden\\  
$^2$Department of Physics and Astronomy, University College London, Gower Street, London WC1E 6BT, UK\\  
$^3$Department of Physics, University of Surrey, Guildford, GU2 7XH, United Kingdom\\  
$^4$ Univ Lyon, Univ Lyon1, Ens de Lyon, CNRS, Centre de Recherche Astrophysique de Lyon UMR5574, F-69230, Saint-Genis- Laval, France \\
$^5$Institute for Computational Science, University of Z\"urich, Winterthurerstrasse 190, CH-8057 Z\"urich, Switzerland
}
              
\date{\today}
\begin{document}
\maketitle

\begin{abstract}
We introduce the ``Engineering Dwarfs at Galaxy formation's Edge'' (EDGE) project to study the cosmological formation and evolution of the smallest galaxies in the Universe. In this first paper, we explore the effects of resolution and sub-grid physics on a single low mass halo ($M_{\rm halo}=10^{9}\Msol$), simulated to redshift $z=0$ at a mass and spatial resolution of $\sim 20\Msol$ and $\sim 3$\,pc. We consider different star formation prescriptions, supernova feedback strengths and on-the-fly radiative transfer (RT). We show that RT changes the mode of galactic self-regulation at this halo mass, suppressing star formation by causing the interstellar and circumgalactic gas to remain predominantly warm ($\sim 10^4$\,K) even before cosmic reionisation. By contrast, without RT, star formation regulation occurs only through starbursts and their associated vigorous galactic outflows. In spite of this  difference, the entire simulation suite (with the exception of models without any feedback) matches observed dwarf galaxy sizes, velocity dispersions, $V$-band magnitudes and dynamical mass-to-light-ratios. This is because such structural scaling relations are predominantly set by the host dark matter halo, with the remaining model-to-model variation being smaller than the observational scatter. We find that only the stellar mass-metallicity relation differentiates the galaxy formation models. Explosive feedback ejects more metals from the dwarf, leading to a lower metallicity at a fixed stellar mass. We conclude that the stellar mass-metallicity relation of the very smallest galaxies provides a unique constraint on galaxy formation physics.
\end{abstract}


\section{Introduction}
\label{sect:intro}
In our current $\Lambda$ Cold Dark Matter ($\Lambda$CDM) cosmological paradigm \citep[e.g][]{komatsu_etal11,planck2014}, galaxies form through the successive `hierarchical' mergers of smaller galaxies \citep[e.g.][]{WhiteRees78,FallEfstathiou80}. This theory has been tremendously successful at matching the observed structure in the Universe on large scales \citep[e.g.][]{Springel2006,vikhlinin_etal09b}. However, on smaller scales there have been long-standing tensions \citep[e.g.][]{2017ARA&A..55..343B}. In part, such tensions owe to the uncertain mapping between stars and dark matter in low mass dwarf galaxies. This arises because dwarfs are very sensitive to the physics of galaxy formation. Isolated dwarfs can have their star formation suppressed or even extinguished by supernovae (SN) feedback \citep[e.g.][]{1986ApJ...303...39D,2000MNRAS.317..697E}, reionisation \citep[e.g.][]{1992MNRAS.256P..43E,1999ApJ...523...54B,2000ApJ...539..517B,2002MNRAS.333..177B}, stellar winds and radiative feedback \citep[e.g.][]{Agertz2013,Hopkins2014}, or some combination of these \citep[e.g.][]{2004ApJ...609..482K,2006MNRAS.371..885R}. If such dwarfs fall into a larger host galaxy, becoming satellites, then they are further buffeted by ram pressure stripping of their interstellar medium \citep[e.g.][]{2003AJ....125.1926G,2013MNRAS.433.2749G} and tides \citep[e.g.][]{2004ApJ...609..482K,2006MNRAS.369.1021M,2006MNRAS.367..387R}.

The sensitivity of dwarfs to galaxy formation physics, like the gas density at which stars form \citep[e.g.][]{2003ApJ...590L...1K,2008PASJ...60..667S} or the details of galactic outflows \citep[e.g.][]{2006MNRAS.371..885R}, makes them a natural `rosetta stone' for constraining galaxy formation models. Early work simulating dwarf galaxies focussed on high resolution small box simulations, stopping at high redshift ($z \sim 5-10$) to avoid gravitational collapse on the scale of the box \citep[e.g.][]{1998ApJ...508..518A,2006MNRAS.371..885R,2006ApJ...645.1054G,2008Sci...319..174M}. These simulations demonstrated that stellar winds, supernova feedback and ionising radiation combine to prevent star formation in halo masses below $\sim 10^{7-8}$\,M$_\odot$ \citep{2006MNRAS.371..885R,2015ApJ...807..154B}. Furthermore, once cooling below $10^4$\,K is permitted and gas is allowed to reach high densities $n_{\rm max} \simgt 10~{\rm cm}^{-3}$ (requiring a spatial and mass resolution better than $\Delta x \simlt 100$\,pc and $m_{\rm min} < 10^3$\,M$_\odot$; \citealt{2012MNRAS.421.3464P,2018arXiv181110625D,2018arXiv181004186B,2018arXiv181003635B}), star formation becomes much more stochastic and violent \citep{2008Sci...319..174M,2012MNRAS.421.3464P,2018arXiv181110625D}. The repeated action of gas cooling and blow-out due to feedback expels dark matter from the galaxy centre, transforming an initially dense dark matter cusp to a core \citep[e.g.][]{2005MNRAS.356..107R,2012MNRAS.421.3464P}. Finally, independent of any internal sources of feedback energy, cosmic reionisation can halt star formation in low mass galaxies \citep[e.g.][]{2006ApJ...645.1054G}. Dwarfs that have not reached a mass of $M_{200} \simgt 10^8$\,M$_\odot$ (see Section \ref{sect:ICs} for a definition of $M_{200}$) by the redshift that reionisation begins ($z \sim 8-10$; \citealt{2014ApJ...793...30G,2018arXiv181111192O}) are gradually starved of fresh cold gas, causing their star formation to shut down by a redshift of $z \sim 4$ \citep{Onorbe2015}. This is similar to the age of nearby `ultra-faint' dwarf galaxies (UFDs) that have $M_* \simlt 10^5$\,M$_\odot$, suggesting that at least some of these are likely to be relics from reionisation, inhabiting pre-infall halo masses in the range $M_{\rm 200} \sim 10^{8-9}$\,M$_\odot$ \citep{2006ApJ...645.1054G,2009ApJ...693.1859B,2011ApJ...741...18B,2014ApJ...789..148W,2014ApJ...796...91B,2018MNRAS.473.2060J,Read2018}. 

To compare quantitatively to such very small galaxies in the local Universe  \citep[e.g.][]{Simon2019}, we need simulations with sufficiently high resolution to model the interstellar medium accurately. Most recent work in this area has focused on isolated dwarfs, removing the need to simultaneously capture a large host galaxy like the Milky Way. While much work has recently been done on simulating the smallest dwarf galaxies, the results from different groups, each of whom make different choices for their sub-grid physics model, box size, hydrodynamic and gravity solver, and resolution, are highly discrepant below $M_{\rm 200} \sim 10^{10}$\,M$_\odot$ \citep[][]{Onorbe2015,Wheeler2015,Sawala2015,Read2016cores,Read2017,Fitts2017,Munshi2017,Maccio2017,Munshi2018,Revaz2018,Wheeler2018,SmithSijacki2019}. While many simulations reproduce the observed structural properties of dwarf galaxies, such as galaxy sizes and velocity dispersions, there are orders of magnitude differences in the stellar mass to halo mass relation, and simulations struggle to reproduce the stellar mass-metallicity relation of the faintest dwarfs (see discussion below; and for a data compilation, see Figures \ref{fig:mstarmhalo} and \ref{fig:MZ}). Furthermore, none of the simulations has yet managed to produce a fully convincing explanation for the existence of \emph{star forming} ultra faint dwarf galaxies like Leo T (except possibly \citealt{Wright2018}, see also \citealt{Verbeke2015}), as cosmic reionisation ought to have evaporated all cold gas from such low mass galaxies. Finally, while observationally there is mounting evidence for dark matter cores in at least some ultra-faint dwarf galaxies \citep{Amorisco2017,Contenta2018,Sanders2018}, some groups find core formation can occur `all the way down' to the very lowest mass dwarfs \citep{Read2016cores,Munshi2017}, while others find that core formation ceases below $\sim 10^{10}$\,M$_\odot$ \citep{Onorbe2015,Tollet2016}.

In this paper, we introduce a new dwarf galaxy simulation campaign: ``Engineering Dwarfs at Galaxy formation's Edge'' (EDGE) with the goal of shedding light on the above discrepancies. There are several new elements to EDGE that make this investigation possible. Firstly, we work at a mass and spatial resolution of $\sim 20\,\Msol$ and $\sim 3$\,pc, respectively, allowing us to better capture the impact of each individual supernova explosion on the forming dwarf. This simplifies the sub-grid modelling of the SN, reducing the need for delayed cooling, momentum capturing schemes, or other similar prescriptions that are required at lower resolution to prevent over-cooling of the SN-heated gas (see e.g. \citealt{2006MNRAS.373.1074S,2008MNRAS.387.1431D,2014MNRAS.438.1985T,2015MNRAS.450.1937C} and for a discussion, see \citealt{Read2016cores}). Secondly, we explore the effect of switching on each piece of our sub-grid model for star formation and feedback one at a time, allowing us to assess its role in shaping the final observed properties of the dwarf today. Thirdly, we model isolated dwarfs in a void region down to redshift zero, using a zoom technique. This allows us to explicitly compare our results with observations of ultra-faint dwarfs that can only been seen, at present, in the Local Group. Finally, we set up our cosmological initial conditions using the new {\sc genetIC} code \citep{Roth16,Rey18}. This will allow us (in forthcoming papers) to forensically explore the effect of different merger histories and environments on the properties of dwarf galaxies. In this first paper in the series, we present the results of 16 hydrodynamical simulations of a single $M_{\rm 200}=10^{9}\Msol$ dwarf run at different resolutions and with different sub-grid physics models. In particular, we explore different star formation prescriptions, the effect of SN feedback and the effect of on-the-fly radiative transfer (RT). We emphasise that our goal in running this large suite of simulations is not to `calibrate' our sub-grid physics model. Rather, we seek to address the following questions: (i) which observables are most sensitive to changes in our sub-grid model? and (ii) which physics are most important for regulating star formation and determining the final observed properties of the smallest dwarfs? In further papers in the series, we will explore the role of physics not considered in this work, the role of different merger histories, the effect of increasing halo mass, and the impact of star formation and feedback on the inner density of the dwarf's dark matter halo.

This paper is organised as follows. In Section \ref{sect:simulations}, we describe the code that we use to run the simulations, the sub-grid physics models that we explore in this work, and how we set up the initial conditions. In Section \ref{sect:results} we present mass growth histories, the stellar mass-halo mass relation and scaling relations for all simulated dwarf galaxies. In Section \ref{sect:discussion}, we compare our findings to observations and previous simulations in the literature and outline the limitations of our models. Finally, in Section \ref{sect:conclusions}, we present our conclusions.

\section{Simulations}
\label{sect:simulations}
We use \ramsesrt{} \citep{Rosdahl2013, Rosdahl2015}, which is an radiation hydrodynamics (RHD)
extension of the cosmological Adaptive Mesh Refinement (AMR) hydrodynamical code \ramses{}
\citep{teyssier02}\footnote{The public code, including all the RHD
  extensions used here, can be downloaded at:
  \url{https://bitbucket.org/rteyssie/ramses}.}, to solve the
evolution of dark matter (DM), stellar populations,
and gas via gravity, hydrodynamics, radiative transfer, and
non-equilibrium radiative cooling/heating. For hydrodynamics, we use the HLLC Riemann solver
\citep{Toro1994} and the MinMod slope limiter to construct gas
variables at cell interfaces from their cell-centred values. To close
the relation between gas pressure and internal energy, we use an ideal gas equation of state with an adiabatic index $\gamma=5/3$. The dynamics of collisionless DM and star particles
are evolved with a multi grid particle-mesh solver and cloud-in-cell
interpolation \citep{GuilletTeyssier2011}. The advection of radiation between
cells is solved with a first-order moment method, using the fully
local M1 closure for the Eddington tensor \citep{Levermore1984} and
the Global-Lax-Friedrich flux function for constructing the inter-cell
radiation field. \report{With the M1 closure, the collisionless nature of photons is lost and beams are not perfectly maintained \citep[see e.g.][]{Rosdahl2013}. This has the effect that in the case where radiation from many sources is mixed, the radiation flux becomes distorted, by up to a factor two \citep[][]{Decataldo2019}, compared to the real flux which can for example be obtained with ray-tracing methods. These more accurate methods, however, are prohibitively expensive to use in simulations with the number of resolution elements and, more importantly, number of radiation sources that we model in this work. We argue that the modest factor of two errors produced by the M1 closure are likely less significant than several other approximations going into our galaxy formation models.
}

In the following sub-sections, we describe the set-up of our
simulations (multi-frequency radiation, thermochemistry, initial conditions) and the adopted galaxy formation physics (star formation and stellar feedback).

\subsection{Radiation}
\label{sect:RT}
Star particles (see \ref{sect:galformphysics}) are treated as single stellar populations (SSPs) with spectral energy distributions (SEDs) taken from \citet{BC03} \citep[see][for details]{Rosdahl2013}. We employ six photon groups to account for 1) photoionisation heating, where three groups bracket the ionization energies for \hi, \hei, and \heii, 2) H$_2$ dissociating Lyman-Werner radiation, 3) direct (single scattering) radiation pressure\footnote{adopting the ``reduced flux approximation'' decribed in Appendix B of \citet{Rosdahl2015b}.}, and 4) non-thermal radiation pressure from multi-scattered IR photons. For dust opacities, we adopt $\kappa=10~(Z/Z_\odot)~{\rm cm^2/g}$ for the IR photons, while for the higher energy photons we assume $\kappa=1000~(Z/Z_\odot)~{\rm cm^2/g}$, with $Z/Z_\odot$ being the gas metallicity in units of the solar value (here taken to be $Z_\odot=0.02$). For a full description of our treatment of dust physics, see \citealt{Rosdahl2015} and \citealt{Kimm2017}. \report{Although we include a treatment of radiation pressure in our RT scheme we find that it has negligible impact, likely due to the low dust content in the simulated UFDs; final galaxy stellar masses in two tests without any radiation pressure (not presented here for brevity) end up within $\pm 20 \%$ of stellar masses in models including it \citep[but see][]{Wise2012}}.

The group properties (average energies and cross sections to molecular hydrogen, hydrogen and helium) are updated every 10 coarse time-steps from luminosity-weighted averages of the spectra of all stellar populations in the simulation volume, as described in \citet{Rosdahl2013}. We do this so that at any time, the cross sections and photon energies are representative of the luminous stellar populations.

In Table~\ref{table:RT} we summarize the above information, and present the average group energies and cross sections, over the entire 13.7 Gyr simulation time from one of our high resolution simulations discussed below\footnote{Similar values are obtained from all of our simulations where RT is included.}. The values only change by a few tens of percent over the course of a simulation as the contributions from both old and metal-rich stellar populations increase. 

\subsection{Gas thermochemistry}
\label{sect:thermochemistry}
For $T>10^4$ K, the contribution from metals to gas cooling is computed using
tables generated with \cloudy{} \citep[][version 6.02]{Ferland1998}. We adopt the UV background of \citet{FG2009} as our standard setting, but also consider that of \citet{haardtmadau96} (see Section ~\ref{sect:simsuite}). \report{The homogenous UV background is necessary to include as we are not capturing radiation sources outside the zoom region. It is not treated using radiative transfer but in the commonly adopted optically thin, cell-by-cell heating approximation \citep[e.g.][]{Katz1996}. Following  \citet{Rosdahl2012} self-shielding against the homogenous UV background is modelled by applying a damping function to the photoionisation rate, $\Gamma_{\rm ss} = \Gamma_{\rm UV}\exp(-n_{\rm H}/10^{-2}~{\rm cm^{-3}})$, for hydrogen densities $n_{\rm H}>10^{-2}~{\rm cm^{-3}}$. Self-shielding against radiation modeled using RT is treated self-consistently.} 

For $T \le 10^4$ K, we use the fine structure cooling rates from \citet{rosenbregman95}. The non-equilibrium hydrogen and helium thermochemistry, coupled with the local ionising radiation, is performed with the quasi-implicit method described in \citet{Rosdahl2013} via photoionisation, collisional ionisation, collisional excitation, recombination, bremsstrahlung, homogeneous Compton cooling/heating off the cosmic-microwave background, and di-electronic recombination. We account for the formation, advection, destruction and cooling of molecular hydrogen (H$_2$) \citep[see][for details]{Nickerson2018}, and its coupling to the radiation field, in all simulations unless otherwise stated.

Along with the temperature, and photon fluxes, we track, in every cell, the fractions of neutral hydrogen, ionised hydrogen, singly, and doubly ionised helium ($x_{\rm H\textsc{i}}$, $x_{\rm H\textsc{ii}}$, $x_\heii$, $x_\heiii$, respectively), and advect them with the gas as passive scalars. The thermochemistry is operator split from the advection of gas and radiation and performed with adaptive time step sub-cycling on every RT time-step. To keep the computational costs low, we use a reduced speed of light, $\bar{c}=10^{-2}c$ \citep[][]{Rosdahl2013}, where c is the true speed of light. 

\begin{table*}
  \begin{center}
  \caption
  {Photon group energy (frequency) intervals and properties from the `Hires+RT' simulation. The
    energy intervals defined by the groups are indicated in units of
    eV by $\epsilon_0$ and $\epsilon_1$. The last four columns show photon
    properties derived every $10$ coarse time-steps from the stellar
    luminosity weighted SED model. These properties evolve over time
    as the stellar populations age, and the mean values are quoted. $\eavg$ denotes the photon
    energies, while $\csavg_{\rm H_2}$, $\csavg_{\hi}$, $\csavg_{\hei}$, and
    $\csavg_{\heii}$ denote the cross sections for ionisation of molecular hydrogen,
    hydrogen and helium, respectively.}
  \label{table:RT}
  \begin{tabular}{llllllllll}
    \hline
    Photon group & \multicolumn{1}{c}{$\epsilon_0$ [eV]} 
    & \multicolumn{1}{c}{$\epsilon_1$ [eV]}
    & $\eavg$ [eV] &  $\sigma_{{\rm H}_2} [\cs]$ & $\csavg_{\hi} \, [\cs]$ 
    & $\csavg_{\hei} \, [\cs]$ & $\csavg_{\heii} \, [\cs]$ \\
    \hline
    IR & 0.1 & 1.0 & 0.6 & 0
            & 0 & 0 & 0 \\
    Optical+FUV & 1.0 & 12.0 & 3.2 & 0
            & 0 & 0 & 0 \\            
     Lyman-Werner & 12.0 & 13.6 & 12.6 & $1.7\times 10^{-19}$
            & 0 & 0 & 0 \\
    UV$_{\hi}$ & 13.6 & 24.59 & 18.1 & $5.1\times 10^{-18}$
            & $3.3 \times 10^{-18}$ & 0 & 0 \\
    UV$_{\hei}$    & 24.59 & 54.42 & 35.6  & $2.0\times 10^{-18}$
            & $5.9 \times 10^{-19}$& $4.3 \times 10^{-18}$ &  0 \\ 
    UV$_{\heii}$   & 54.42  & \multicolumn{1}{c}{$\infty$} 
            & 64.9 & $3.4\times 10^{-19}$ & $8.3 \times 10^{-20}$ & $1.2 \times 10^{-18}$ & 
                                        $1.1 \times 10^{-18}$ \\
    \hline
  \end{tabular}
  \end{center}
\end{table*}

\subsection{Galaxy formation physics}
\label{sect:galformphysics}

Star formation follows a Schmidt law,
\begin{equation}
\label{eq:schmidt}
\dot{\rho}_{*}=\epsilon_{\rm ff}\frac{\rho_{\rm g}}{t_{\rm ff}}\,\,{\rm for} \,\,\rho_{\rm g}>\rho_{\star},
\end{equation}
where $\rho_{\rm g}$ is the gas density, $\rho_\star$ the density threshold of star formation, $t_{\rm ff}=\sqrt{3\pi/32G\rho}$ is the local gas free-fall time and $\epsilon_{\rm ff}$ is the star formation efficiency per free-fall time. We adopt $\rho_{\star}=300~{m_{\rm H}}~\cc$, and sample Eq.~\ref{eq:schmidt} stochastically on a cell-by-cell basis at every fine simulation time step using $300\Msol$ star particles\footnote{Chosen to accommodate the (discrete and stochastic) mass loss from several SN type II explosions, which is not the case if initial star particle masses are set too close to the mass of individual massive stars in our current feedback scheme. Note also that star particle masses,on average, are reduced by up to $50\%$ due to stellar evolution \citep[e.g.][]{Leitner2012}.} \citep[see][]{Agertz2013}. Furthermore, we only allow stars to form from cold gas ($T<100$ K). In a subset of simulations, we require star formation to only occur in molecular gas, as motivated by the observed close to linear relation between $\Sigma_{\rm mol}$ and $\Sigma_{\rm SFR}$ \citep[e.g.][]{bigiel2008}. In those simulations, $\rho_{\rm g}\rightarrow f_{{\rm H}_2}\rho_{\rm g}$ in Eq.~\ref{eq:schmidt}, where $f_{{\rm H}_2}$ is the molecular hydrogen fraction in a cell \citep[see also][]{Gnedin09, GnedinKravtsov2010,Christensen2014}. \report{We note that while a correlation between molecular hydrogen and star formation is well motivated both theoretically and empirically, it is not well established in the low metallicity regimes probed in this work}\footnote{\report{The cooling and star formation time scales can in metal poor environments be shorter than the time scale for reaching an equilibrium chemical state for which the gas would be H$_2$-dominated. Star formation can under such conditions correlate with atomic gas \citep[e.g.][]{Krumholz2012}. The non-equilibrium treatment of H$_2$ in our adopted method \citep[][]{Nickerson2018} mitigates this issue \citep[see also][]{krumholzgnedin2011}.}} \report{\citep[{\rm [Fe/H]}$\lesssim -2$,][]{Glover2012,Krumholz2012}. For this reason we consider the use of H$_2$-based star formation models in the UFD regime as exploratory \citep[see also][]{Munshi2018}.}

Observationally, $\epsilon_{\rm ff}$ averages $1\%$ on galactic kpc scales \citep[][]{bigiel2008} as well as in Milky Way giant molecular clouds (GMCs) \citep[][]{krumholztan07}, albeit with a spread of several dex \citep[][]{Murray2011b,Lee2016}. Recently \citet{Grisdale2019} demonstrated how high efficiencies ($\epsilon_{\rm ff}\sim 10\%$) on scales of parsecs, coupled to a feedback budget like the one adopted here, provides a close match to the observed (i.e. emerging) efficiencies on scales of individual GMCs. Motivated by these findings we adopt $\epsilon_{\rm ff}=10\%$.  

We adopt the stellar feedback budget described in \citet{Agertz2013}. Briefly, this feedback prescription includes the injection of energy, momentum, mass and heavy elements over time from SNII and SNIa explosions and stellar winds into the surrounding ISM. In contrast to  \citet{AgertzKravtsov2015}, we do not adopt a subgrid model for radiation pressure, as this is self-consistently treated by the RT solver (see Section ~\ref{sect:RT}, and Section ~\ref{sect:simsuite} for which simulations adopt RT). Each mechanism depends on the stellar age, mass and gas/stellar metallicity, calibrated on the stellar evolution code STARBURST99 \citep{Leitherer1999}, treating each formed stellar particle as a single-age stellar population with a \citet{Kroupa01} initial mass function (IMF). Feedback is done continuously at the appropriate times when each feedback process is known to operate, taking into account the lifetime of stars of different masses within a stellar population. 

We track iron (Fe) and oxygen (O) abundances separately, and advect them as passive scalars. When computing the gas cooling rate, which is a function of total metallicity, we construct a total metal mass as:
\begin{equation}
M_{Z}=2.09M_{\rm O}+1.06M_{\rm Fe}
\end{equation}
according to the solar abundances of alpha (C, N, O, Ne, Mg, Si, S) and iron (Fe, Ni) group elements of \citet{Asplund2009}.

SNe explosions are modelled as discrete events, and we follow the approach by \citet{KimOstriker2015} \citep[see also][]{Martizzi2015} and inject the full momentum generated during the Sedov-Taylor phase if a supernova cooling radius is not captured with at least 6 grid cells, otherwise we inject a $E_{\rm SN}=10^{51}$ ergs of thermal energy \citep[see][for details]{AgertzRomeoGrisdale2015} and allow for the hydrodynamic solver to track the buildup of momentum. At the numerical resolution adopted here, $>90\%$ of all SN explosions are resolved by at least 6 cooling radii in our fiducial simulation. \report{We note that we do not enforce any additional refinement criterion to achieve this; the (Lagrangian) mass based refinement scheme, discussed in Section \ref{sect:simsuite}, is enough for this to be satisfied.}
This feedback budget has been shown to lead to Milky Way disc galaxies in close agreement to observations \citep[][Agertz et al. in prep]{AgertzKravtsov2016}, bursty star formation and realistic properties of dwarf galaxies \citep[][]{Read2016cores,Read2016galrot}, and an interstellar medium (ISM) and structure of giant molecular clouds in excellent agreement with observations \citep{Grisdale2017, Grisdale2018}.

Finally, in order to account for enrichment from unresolved Population III (Pop III) star formation, we adopt a pre-existing metal floor at $Z=10^{-3}Z_\odot$ \citep[e.g.][]{wise_etal12,Jaacks2018} added to the oxygen field\footnote{Motivated by the abundance ratios for alpha group elements observed in extremely metal poor stars \citep[][]{Iwamoto2005}, but we note that Pop III yields are \emph{extremely} uncertain, and that abundance ratios and the overall metal content is likely dependent on environment \citep[e.g.][]{Jaacks2018}.}. We note however that H$_2$ cooling is the main coolant at $Z\lesssim 10^{-2}Z_\odot$, which is the regime we are exploring in this work. As soon as star formation and enrichment begins, metal line cooling, and subsequently atomic line cooling, also become important coolants \citep{Wise2014}. The inclusion of a Pop III floor therefore has a small effect on star formation properties in our simulations, and tests with $Z\leq 10^{-4}Z_\odot$ show that our choices have no impact on any of the conclusions presented in this paper.

\begin{table*}
	\centering
	\caption{Simulations and their $z=0$ properties. All simulations use the same cosmological zoom initial conditions, are run with {\small RAMSES-RT} and reach a minimum cell size of $\Delta x_{\rm min}=3\pc$. The specified quantities are, from left to right: the dark matter particle mass in the deepest refinement region; the gas refinement mass above which new cells will be opened; the total stellar mass formed by $z=0$; the $V$-band magnitude computed using {\sc sunset} ray-tracing; the half-mass radius; the 1D equivalent velocity dispersion; the dynamical mass-to-light ratio within this radius; the enrichment relative to solar; and the mean star formation rate over the time until the galaxy quenches.
	}
	\label{tab:sims}
	\begin{tabular}{lccccccccc} 
		\hline
		Simulation & $m_{\rm DM}$ &  $m_{\rm bar}$ & $M_{\star}$ & $M_V$ &  $r_{1/2}$& $\sigma_\star$ & $M_{\rm dyn}/L$ & [Fe/H] & $\langle {\rm SFR} \rangle^a$  \\
				 & [$\Msol$] & [$\Msol$] & [$10^5~\Msol$] & & [pc] & [$\kms$] & ($<r_{1/2}$) & [dex] & [$\Msol/$yr]\\
		\hline
		Fiducial,  no feedback & 945 & 161 & 172.1 & -12.0 & 51 & 17.6 & 2 & no enrichment & $1.4\times 10^{-2}$\\
		Fiducial & 945 & 161 & 1.2 & -6.6 & 313 & 6.1 & 287 & $-2.65\pm 0.82$ & $1.1\times 10^{-4}$ \\
		Fiducial (weak feedback)$^b$  & 945 & 161 & 15.8 & -9.4 & 336 & 6.1 & 17 & $-1.07\pm 0.68$ & $1.4\times 10^{-3}$\\
		Fiducial, $2\times E_{\rm SN}$& 945 & 161 & 1.2 & -6.6 & 280 & 6.3 & 212 & $-2.66\pm 0.79$ & $1.1\times 10^{-4}$\\
		Fiducial, $10\times E_{\rm SN}$& 945 & 161 & 0.52 & -5.7 & 314 & 6.9 & 647 & $-3.42\pm 0.96$ & $4.8\times 10^{-5}$\\
		Fiducial, $100\times E_{\rm SN}$& 945 & 161 & 0.11 & -4.0 & 157 & 5.6 & 949 & $-4.00\pm 1.20$ & $2\times 10^{-4}$ \\
		Fiducial, H$_2$ SF & 945 & 161 & 1.2 & -6.6 & 285 & 6.1 & 189 & $-2.49\pm 0.99$ & $1.1\times 10^{-4}$\\
		Fiducial, no H$_2$ physics & 945 & 161 & 0.53 & -5.7 & 265 & 5.6 & 338 & $-2.27 \pm 1.29 $ & $4.8\times 10^{-5}$ \\
		Fiducial, HM UV & 945 & 161 & 0.94 & -6.35 & 283 & 6.6 & 287 & $-2.59\pm 0.79$ & $9.4\times 10^{-5}$\\
		Fiducial + RT & 945 & 161 & 0.30 & -5.7 & 308 & 5.6 & 398 & $-2.31\pm 0.88$ & $2.7\times 10^{-5}$ \\
		Fiducial + RT (weak feedback)$^b$ & 945 & 161 & 1.1 & -7.1 & 412 & 6.5 & 200.8 & $-2.35\pm 0.96$ & $9.7\times 10^{-5}$\\
		Fiducial + RT,  H$_2$-based SF & 945 & 161 & 0.62 & -6.5 & 597 & 6.5 & 535 & $-1.91\pm 0.65$ & $5.3\times 10^{-5}$\\

		\hline	
		Hires & 118 & 20 & 2.5 & -7.4 & 370 & 7.2 & 171 & $-2.61 \pm 0.95$ & $2.3\times 10^{-4}$\  \\
		Hires (weak feedback)$^b$ & 118 & 20 & 20.5 & -9.7 & 311 & 6.1 & 12.3 & $-1.01 \pm 0.91$ & $1.8\times 10^{-3}$ \\
		Hires + RT & 118 & 20 & 0.31 & -5.7 & 203 & 5.2 & 226 & $-2.6\pm 1.37$ & $2.9\times 10^{-5}$ \\
		Hires + RT (weak feedback)$^b$ & 118 & 20 & 2.8 & -8.1 & 368 & 6.3 & 67.1 & $-1.32\pm 0.85$ & $2.3\times 10^{-4}$\\

		\hline
		Fiducial, dark matter only & 1106 & - & - & - & - & - & - \\
		Hires, dark matter only & 138 & - & - & - & - & - & - \\
		\hline
	\end{tabular}
	\begin{flushleft}
	\footnotesize{$^a$ Defined as $M_\star/t_{\rm SF}$, where $t_{\rm SF}$ is the duration of star formation for the galaxy, here $\sim 1-1.2$ Gyr in all simulations}\\
	\footnotesize{$^b$ Maximum allowed gas temperatures $T_{\rm max}=10^8~{\rm K}$, maximum allowed SN and stellar winds velocities $\vel_{\rm fb,max}=10^3\kms$}
\end{flushleft}
	
\end{table*}

\subsection{Initial conditions}
\label{sect:ICs}
In this work we study the cosmological formation of dwarf galaxies forming in isolation, hence removing complexities such as gas stripping during infall and environmental star formation quenching. Being isolated, deformation due to tides is also minimized, and we can study the ``pristine" galaxy formation scenario in a $10^9\Msol$ dark matter halo. To generate initial conditions, we use the code \textsc{genetIC} \citep{Roth16,Rey18} which will in future work allow us to explore a continuum of alternative merger histories for the same galaxy. For the present paper, the most important capability of \textsc{genetIC} is simply to recursively refine regions of the simulation on extremely fine grids. We first generated initial conditions for a simulation with box size of $L_{\rm box}=50~\Mpc$ with cosmological parameters  $\Omega_{\rm m}=0.309$, $\Omega_\Lambda=0.691$, $\Omega_{\rm b}=0.045$ and $H_0=67.77~\kms~\Mpc^{-1}$, in line with data from the {\small PLANCK} satellite \citep{planck2014}. 

We simulated this volume with only dark matter from $z=99$ to $z=0$, using $512^3$ resolution elements (giving a particle mass of $m_{\rm DM}=3.8 \times 10^7 \Msol$). Next, we picked the largest void and resimulated it (again with only dark matter) at the equivalent of $2048^3$ resolution ($m_{\rm DM}=4.9 \times 10^5 \Msol$), adding the appropriate small scale power to this grid. We then identified dark matter haloes within the void at $z=0$ using the HOP halo finder \citep[][]{HOP1998} and computed their virial masses, $M_{200}$, defined as the mass inside of a spherical volume encompassing $200$ times the cosmic critical density $\rho_{\rm crit}=3H(z)^2/8\pi G$, where $H(z)$ is the Hubble parameter. The radial extent of this spherical volume defines the virial radius, $r_{200}$.

To find haloes in isolation, we computed pairwise distances, measured from the edges of their individual virial radii. For halo $n$ and $m$, the distance is $D_{n,m}=|\mathbf{r}_n-\mathbf{r}_m|-r_{200,n}-r_{200,m}$, where $\mathbf{r}$ is the position of a halo center. Expressed in units of the virial radius of halo $n$,  $I_{n,m}=D_{n,m}/r_{200,n}$. As a measure of isolation for halo $n$, we define an \emph{isolation parameter} as the minimum of $I_{n,m}$, i.e. $I_{n}={\rm min}(I_{n,m})$. At $z=0$, we only considered halos in the mass range \rm $0.8\times 10^9\Msol<M_{200}<1.2\times 10^9\Msol$, with $I_{n}>10$ for halo pairs with $M_{200,m}/M_{200,n}>1$. 

From the filtered halo catalogue, we picked a halo that at $z=0$ had a mass of $M_{200} \sim 10^{9}\Msol$ and was visually isolated from massive dark matter filaments. Having found a dark matter halo satisfying our selection criteria, we identified all particles belonging to this halo out to $2\times r_{200}$ at $z=0$ and traced the particles back to the initial conditions ($z=99$). 
For the Lagrangian region of this halo, we generated separate new initial conditions at the equivalent of $16\,384^3$ resolution ($m_{\rm DM}=1106~\Msol$), again adding small scale density fluctuations compatible with the background. These `fiducial' initial conditions are then modified to include baryons as well as dark matter, forming the basis for the simulation suite described below. For the `Hires' simulations, we further refined the Lagrangian region by a factor of 8 in mass. 

\begin{figure*}
	\begin{center}
		\includegraphics[width=0.9\textwidth]{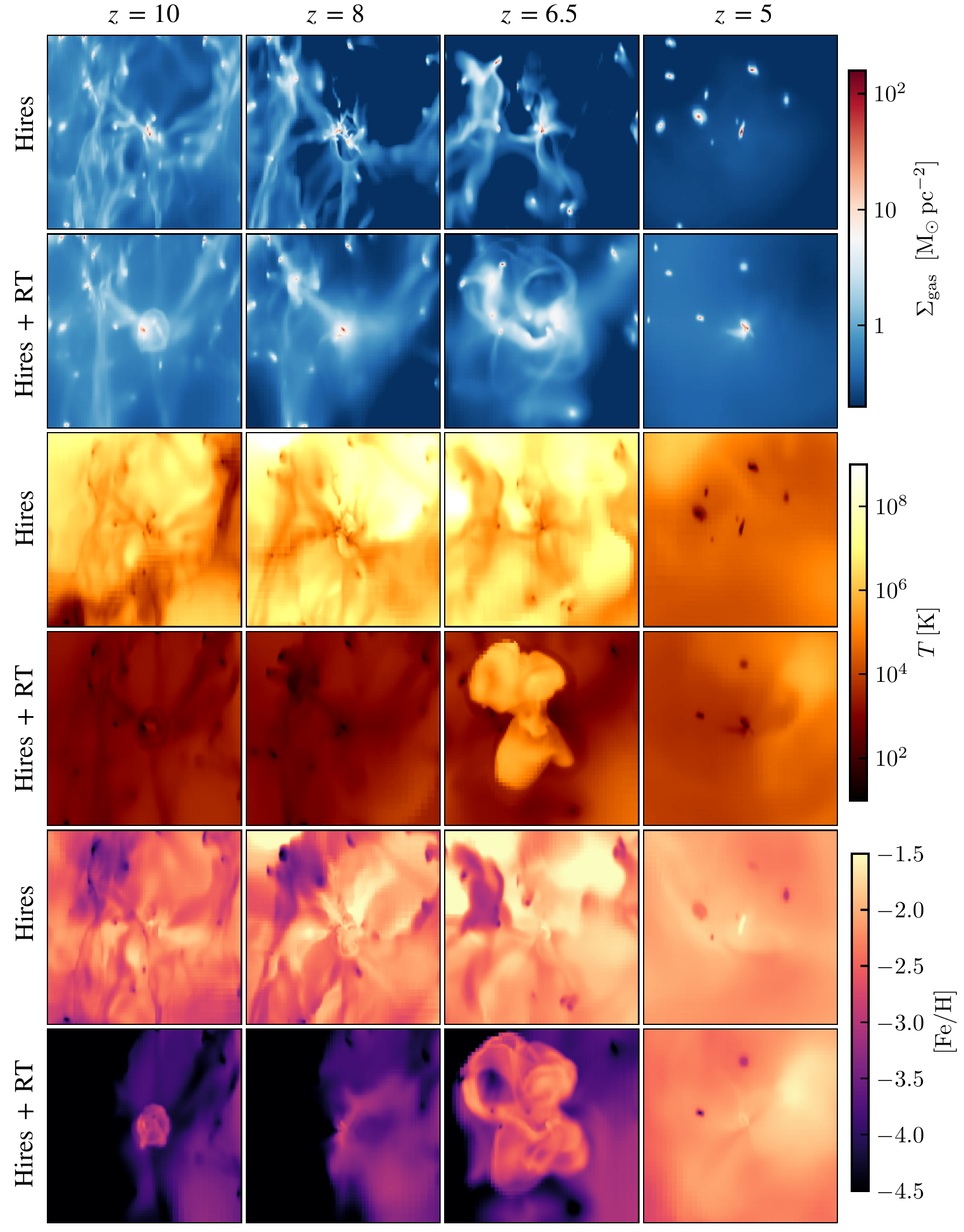}
		\caption{Visual comparison of the gas surface density (top), temperature (middle) and metallicity (bottom) in a $5\times5\kpc^2$ region \report{(proper kiloparsecs)} centered on the main dwarf galaxy in the `Hires'  and `Hires+RT' simulations at $z=10, 8, 6.5$ and $5$. Prior to reionisation ($z\gtrsim 6.5$),
		local radiative feedback in `Hires+RT' completely changes how the galaxy self-regulates; the internal radiation sources are able to keep the gas warm ($T\sim 10^4$ K), leading to reduced star formation and weaker outflows at these early times. Conversely, without RT, large scale SN driven outflows, together with the significantly weaker contribution from stellar winds, are the only regulation mechanism, resulting in a  hot ($T>10^6$ K) and enriched ([Fe/H]$> -3$) circumgalactic medium. At late times, the differences in the intergalactic gas in the two cases diminish because cosmic reionisation (which is included even in the `Hires' simulation) takes over as the dominant radiation source. 
		}
		\label{fig:maps}
	\end{center}
\end{figure*}

\subsection{Simulation suite}
\label{sect:simsuite}

For halos of a given mass, there will be a diversity of cosmologically-determined mass accretion histories and environments giving rise to a spread in final observed properties at $z=0$. Our EDGE project will ultimately explore this diversity  using ``genetic modification" \citep{Roth16,Rey18}. However, before assessing the connection between history and observables, we need a firm handle on theoretical uncertainties resulting from the small-scale star formation and feedback physics. While there will never be a `complete' account of feedback physics, understanding the leading-order uncertainties and their implications for interpreting future simulations is a key first step. 

This first work therefore probes the effect of differing physics implementations, and we undertake this study at two numerical resolutions. In the high resolution simulations, the dark matter particle resolution is $m_{\rm DM}=118\Msol$, and the equivalent baryon resolution (the mass of baryons at the finest grid level in the initial conditions) is $m_{\rm bar}=\Omega_{\rm b}/\Omega_{\rm m}m_{\rm DM}=20\Msol$. For our fiducial resolution, the corresponding numbers are $m_{\rm DM}=945\Msol$ and $m_{\rm bar}=161\Msol$. In all simulations we reach a mean physical resolution of $\Delta x = 3~\pc$ at all times in the inner parts of dark matter haloes and in the ISM. Refinement is based on a pseudo-Lagrangian approach, where a cell is split if its mass $m_{\rm cell}$ exceeds $8\times m_{\rm bar}$, where the cell mass accounts for both stars and gas. In addition, a cell is allowed to refine if it contains more than 8 dark matter particles. All simulations are run from $a=0.01$ ($z=99$) to $a=1$ ($z=0$), and simulation snapshots are stored every $\Delta a=0.01$.

For both resolution settings, we adopt the standard galaxy formation physics presented above and refer to these as `Fiducial+RT' and `Hires+RT', and `Fiducial' and `Hires' when RT is not included. We reemphasise that without on-the-fly RT, we do not consider any \emph{subgrid} model of radiative feedback \citep[e.g.][]{Agertz2013}, only feedback from stellar winds and supernovae. For the fiducial resolution setting, we carry out a suite of simulations where the sensitivity to galaxy formation physics is tested as follows: increasing the strength of SN feedback ($E_{\rm SN}=2, 10$ and $100 \times 10^{51}$), not considering H$_2$ physics and the associated gas cooling, changing the UV background field (changing from the fiducial \citealt{FG2009} UV background to that of \citealt{haardtmadau96}), adopting an H$_2$ based star formation model (see Section ~\ref{sect:galformphysics}) and studying the impact of coupling it to RT. In the latter model, molecular hydrogen can be dissociated by Lyman-Werner radiation, forcing stars to only form in environments where H$_2$ is self-shielded, as well as shielded by dust.

For both resolution settings, with and without RT, we also carry out simulations where we artificially limit the efficiency of SN wind driving by limiting the maximum allowed temperatures of the gas to $T_{\rm max}=10^8$ K and restrict the maximum allowed velocities of feedback, upon injection, to $\vel_{\rm fb,max}=1000~\kms$ (approximately leading to post-shock temperatures of $\sim 10^8$ K)\footnote{In the adopted feedback model, supernova ejecta from $8~\Msol$ stars travel with velocities of $\sim 3000~\kms$}. We refer to these models as `weak feedback'. Although the exact values of such feedback limiters are arbitrary, these tests illustrate the role of hot, fast-moving winds in regulating ultra-faint dwarf formation, as we will demonstrate below. We summarise the entire simulations suite, as well as the $z=0$ properties of the central dwarf galaxy, in Table~\ref{tab:sims}.

\begin{figure*}
	\begin{center}
		\includegraphics[width=0.95\textwidth]{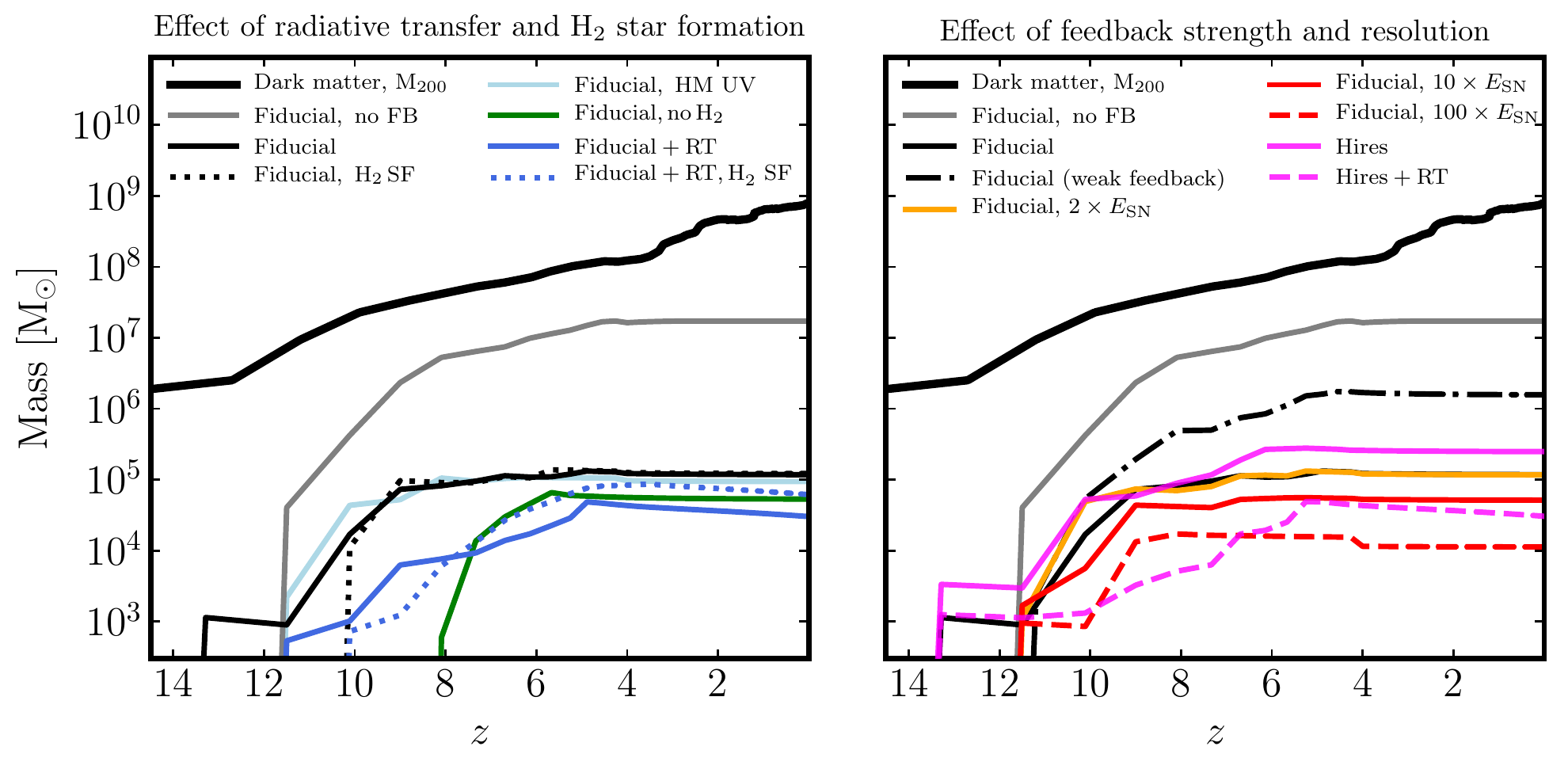} 
		\caption{Growth histories of dark matter and stars from our simulation suite presented in Table \ref{tab:sims}. Without feedback (`Fiducial, no FB'), stars form efficiently until $z\sim 4$, with a final stellar mass $M_\star>10^7\Msol$. Supernova feedback brings the mass to $\sim10^5\Msol$ (`Fiducial') or $\sim 2.5\times 10^5\Msol$ (`Hires'), with radiative feedback suppressing star formation even further ($\sim 3\times 10^4\Msol$, `Fiducial + RT' and `Hires + RT'). Changes to other galaxy formation physics, such as the UV background, neglecting H$_2$ cooling, star formation based on H$_2$, can modify the early phases of star formation, but all results in final stellar masses within factor of $\sim 2-3$ of each other.
		}
		
		\label{fig:growth}
	\end{center}
\end{figure*}

\section{Results}
\label{sect:results}
Figure \ref{fig:maps} shows the gas surface density, temperature and metallicity in a $5\times5\kpc^2$ region centered on the main dwarf galaxy in the `Hires' and `Hires+RT' simulations at $z=10, 8, 6.5$ and $5$. The maps illustrate the dramatic effect of stellar feedback, and how the inclusion of radiative feedback changes the mode of galaxy formation. In `Hires', star formation leads to large-scale hot ($T>10^6\,{\rm K}$) SN driven outflows at all times before $z\sim5$. The enriched winds efficiently mix into the intergalactic medium (IGM), leading to an average background [Fe/H]$> -3$ even at early times ($z\sim 10$). 

This is in stark contrast to the RT counterpart, where the gas is kept warm ($T\sim 10^4$ K) and large scale metal rich outflows are relatively weak until $z\lesssim 6.5$. Before this epoch, the IGM is metal poor ([Fe/H]$\lesssim -4$), and metal line cooling is subdominant to H$_2$ cooling. At $z=5$, reionisation has been operating for $\sim 500$ Myr, and large scale filamentary structures have been evaporated, with stars still forming from the residual cold gas from earlier accretion epochs. At this time, both simulations feature a circumgalactic medium enriched to [Fe/H]$\sim-2$. `Hires', due to its early intense outflows, retains an enriched intergalactic medium to larger radii than `Hires + RT', and the RT simulation also features a visibly denser circum-galactic medium compared to 'Hires'. In the next sections we identify differences in the interstellar medium that give rise to the modified outflow behaviours.

\subsection{Mass growth histories}
\label{sect:growth}
Figure~\ref{fig:growth} shows the buildup of stellar and dark matter halo masses ($M_{200}$) for the simulations in Table~\ref{tab:sims}. Stellar masses ($M_\star$) are throughout this paper defined as the total mass of stars in the inner part of the dark matter halo ($r<0.25R_{200}$). The left hand panel includes the models targeting radiative transfer and H$_2$ physics, while the right panel focuses on variations in feedback strength and numerical resolution. Star formation starts at $z\sim 12-13$ in all models, with delays introduced by variations in how H$_2$ physics is treated; see below. The virial temperature of the halo at $z<10$ is $T_{\rm vir}\sim 5000-9000$ K, i.e. below the `atomic-cooling' regime ($T_{\rm vir}\gtrsim 10^4$ K). As reionisation heats the gas to $T\sim 10^4$ K, this leads to a cessation of gas accretion and ultimately quenching of star formation. Star formation from the already cold and dense (self-shielded) gas in the ISM can, however, proceed throughout the epoch of reionisation. Indeed, the last stars form around $z\sim 4-5$ in all simulations.

Without any stellar feedback, the final stellar mass is close to $M_\star=2\times10^7\Msol$. By including stellar feedback (but neglecting RT), stellar masses are lowered by over two orders of magnitude to $M_\star = 1.2\times 10^5\Msol$ in `Fiducial', and $\sim 2.5\times 10^5 \Msol$ in `Hires'. This result is independent of the adopted UV background, with `Fiducial, HM UV' and `Fiducial' having very similar mass growth histories. We next turn to how adopted galaxy formation physics affects stellar masses using our fiducial simulations suite.

\subsubsection{Impact of molecular hydrogen}
All of our simulations include H$_2$ formation, destruction and cooling (see Section \ref{sect:thermochemistry}), with the exception of the test run `Fiducial, no H$_2$ physics'. However typically we determine star formation based on local density and temperature of all gas (Section \ref{sect:galformphysics}). In one variant (`Fiducial, H$_2$ SF') we tie the star formation exclusively to the molecular component. These two variants thus allow us to determine the overall importance of molecules in our formulation.

In order to reach sufficiently high H$_2$ gas fractions for star formation, H$_2$ needs to be self-shielded. In the low metallicity ISM of the simulated ultra-faint dwarfs (see Section \ref{sect:scalings}), this requires high densities, and we find that star formation occurs only at densities $n\sim 10^3-10^4 ~{\rm cm}^{-3}$ in `Fiducial, H$_2$ SF'. As a result, the onset of star formation is slightly delayed ($z\sim 10$) compared to the fiducial model; nonetheless, as soon as star formation starts, the stellar mass growth quickly catches up with the `Fiducial' galaxy. Overall, therefore, the star formation law is not strongly sensitive to molecules in itself. This insensitivity may not comes as a surprise as star formation anyway is restricted to dense ($n>300~{\rm cm}^{-3}$) and cold gas ($T<100$~K), which benefits molecular hydrogen formation \citep[see also][]{Hopkins2018}.

On the other hand, completely neglecting H$_2$ cooling (`Fiducial, no H$_2$ physics') delays the onset of star formation more significantly (to $z \sim 8$); the final stellar mass is then suppressed by a factor of two relative to `Fiducial'. This indicates that the onset of star formation in cold dense gas is sensitive to details of the cooling function. Once star formation starts, enrichment and metal line cooling allows approximately the same gas reservoir to be available for star formation in all models after $z\simeq 8$, limiting the impact on the final stellar content. We conclude that molecular hydrogen physics is a subdominant uncertainty relative to other factors that we explore below.

\begin{figure*}
	\begin{center}
		\includegraphics[width=0.93\textwidth]{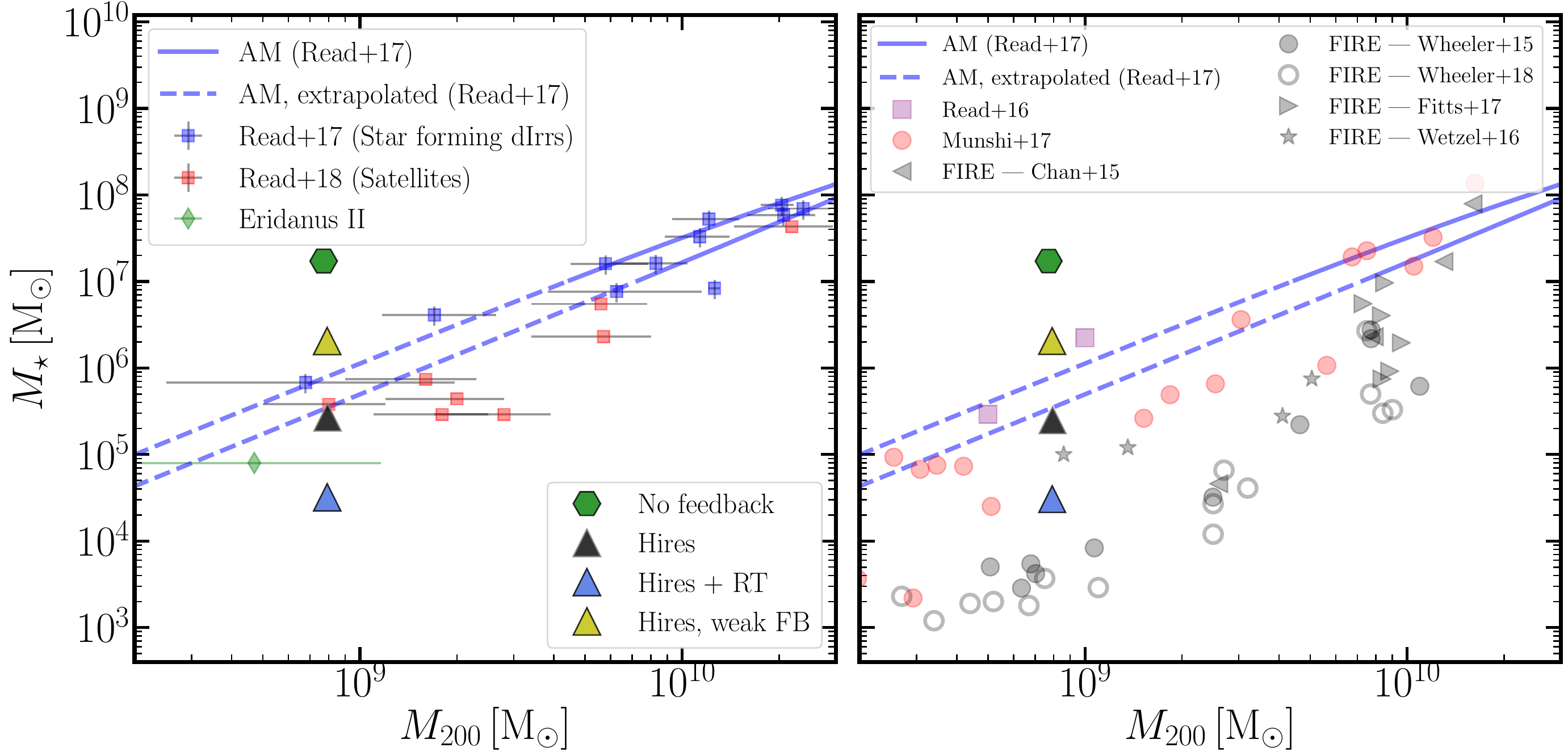} 
		\caption{Stellar mass -- halo mass relation for our high resolution  simulation suite at $z=0$, compared (left) to local group dwarfs and (right) to existing simulation suites. Abundance matching (AM) extrapolation results from \citet{Read2017} are overplotted on both panels. In the left panel, blue squares are irregular dwarfs taken from \citet{Read2017} and red squares are data for (non-star forming) satellite dwarfs compiled by \citet{Read2018}; the latter are probably the most appropriate point of comparison since our simulated object is always quenched by reionisation. A purely SN driven scenario (`Hires') suppresses star formation by 2 orders of magnitude compared to the model neglecting feedback, and results in a galaxy formation efficiency ($M_\star/M_{200}$) in line with AM. Radiative feedback lowers $M_\star/M_{200}$ by almost an additional order of magnitude, within the uncertainties of UFDs such as Eridanus II. In the right panel, the grey points show results from different incarnations of the FIRE project; red circles give the relation found by \citet{Munshi2017}; and pink squares show the isolated dwarf galaxy simulations of \citet{Read2016cores}. }
		\label{fig:mstarmhalo}
	\end{center}
\end{figure*}

\begin{figure*}
		\includegraphics[width=0.93\textwidth]{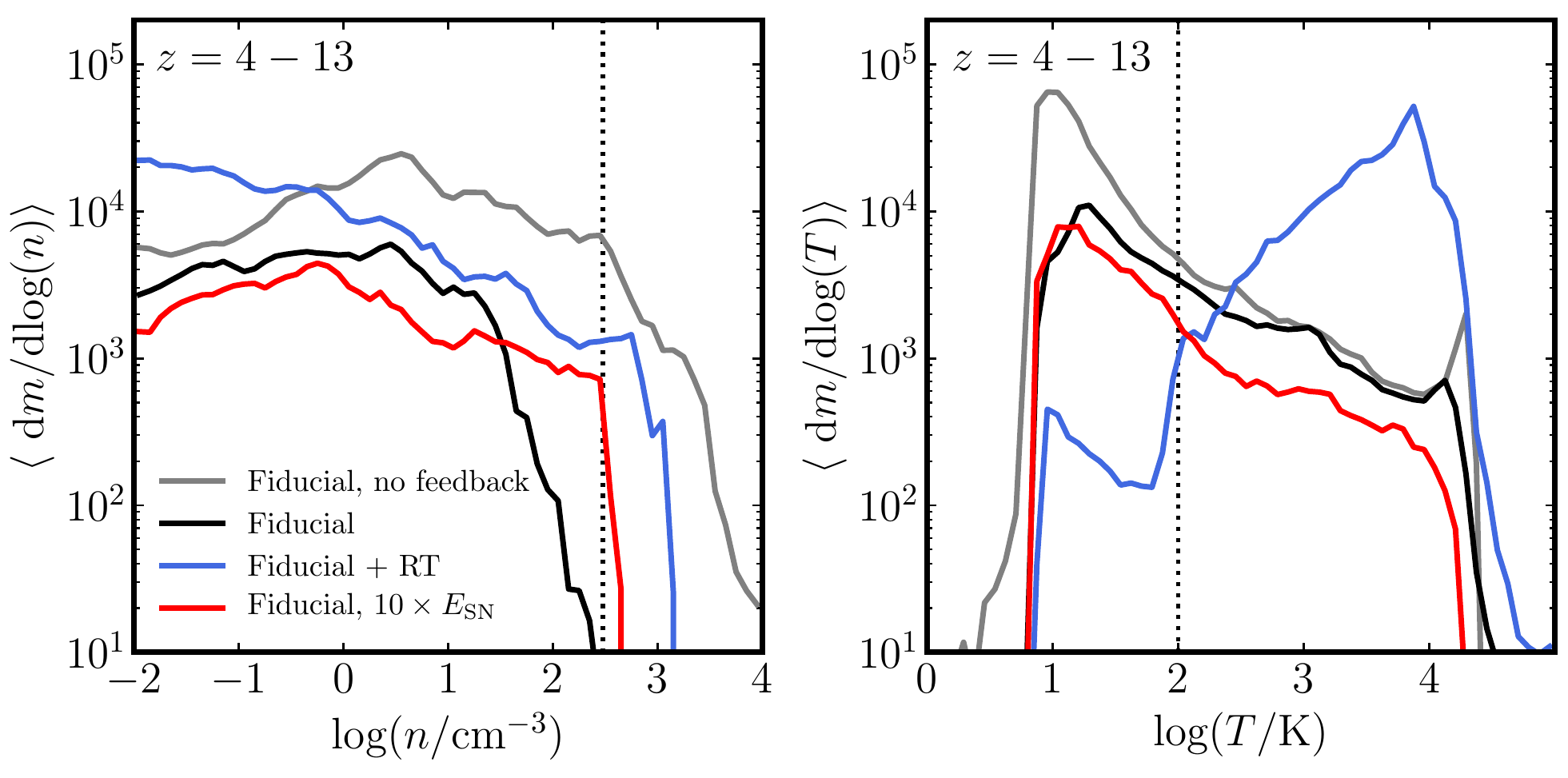} 
		\caption{(Left) Mass weighted density PDFs at $z=4-13$. (Right) Mass weighted temperature PDFs at $z=4-13$. Grey, black, blue and red lines indicate no-feedback, `Fiducial', `Fiducial+RT' and `Fiducial $10\times E_{\mathrm{SN}}$' cases. Dotted vertical lines indicate the adopted density ($n=300~{\rm cm}^{-3}$) and temperature thresholds ($T=100$ K) for star formation. The supernova driven models regulate star formation mainly by blowing out gas from the galaxy, limiting the amount of dense star forming gas. In contrast, radiative feedback lowers star formation by maintaining most of the ISM in a warm ($T\sim 10^4$ K) non-star forming phase, with less vigorous outflows and more gas remaining in the inner halo at all times.
		}
		\label{fig:gasevol}
\end{figure*}

\subsubsection{Impact of supernova feedback strength}
Increasing the available energy per SN explosion by a factor of two (`Fiducial, $2\times E_{\rm SN}$') has almost no effect on the stellar mass growth histories. However, an increase by a factor 10 (or 100) results in a suppression by a factor of 2 (or 10) in final stellar masses. This reveals a weak, sub-linear dependence of the amount of star formation on the strength of SN feedback, compatible with results for simulated disc galaxies \citep[][]{Benincasa2016}. 

Limiting the effect of stellar feedback (`Fiducial, weak feedback'), by artificially imposing numerical ceilings on the allowed supernova gas temperature $T_{\mathrm{max}} = 10^8\,\mathrm{K}$ and velocities ($v_{\mathrm{fb,max}} = 10^3\,\mathrm{km\,s^{-1}}$), results in an increased stellar mass by an order of magnitude. This stems from the fact that a large fraction of supernova explosions occur in low density gas where SN bubble temperatures and velocities exceed the ceilings (see Section \ref{sect:simsuite}). Capturing this extreme temperature gas phase, and the associated fast ejecta velocities, is thus essential to capture the full effects of energetic feedback at high resolution.

\subsubsection{Impact of radiative feedback}
As previously discussed with reference to Figure~\ref{fig:maps}, including radiative feedback (`RT' simulations) generates a major shift in the early behaviour of our galaxy. The `Fiducial' and `Hires' models with RT all result in a final galaxy with stellar masses $M_\star\approx 3 \times 10^4\Msol$ -- an additional factor of $\sim 5-10$ reduction from pure SN regulation, and the lowest stellar masses recovered in our simulation suite. Radiative feedback coupled to H$_2$ based star formation, i.e. allowing for H$_2$ destruction by Lyman-Werner radiation from young stars (`Fiducial + RT,  H$_2$-based SF'), particularly suppresses star formation rates at early times ($z\gtrsim 6$), although final stellar masses are close to that of `Fiducial + RT'.
	
 Without feedback, star formation rates (averaged in age bins of width $100$ Myr) reach $\dot{M_\star}\sim 10^{-2}\msunyr$ at a lookback time of $t_{\rm lookback}\sim 12.5-13\Gyr$. The order of magnitude change in final stellar mass when feedback processes are introduced are mirrored in the average star formation rates, with `Fiducial' and `Fiducial+RT' models peaking at $\sim 10^{-4}\msunyr$ and $\sim {\rm few}\times 10^{-5}\msunyr$ respectively, when averaged over $100$ Myr windows. 
 
 It is worth noting that this effect is far greater than the differences generated by changing numerical resolution.  `Hires + RT' forms a total of $3.1 \times 10^4\,\Msol$ compared to $3.0 \times 10^4$ for `Fiducial + RT'. Enabling RT actually seems to minimise resolution sensitivity, most likely because more gas is kept in a warm and relatively diffuse phase. 

\subsection{The stellar mass -- halo mass relation}
\label{mstarmhalo}
In Figure~\ref{fig:mstarmhalo}, we show the relation between the galaxy stellar mass and dark matter halo mass (the $M_\star-M_{200}$ relation) for our simulations. For clarity, we show results from the `Hires' models; the corresponding `Fiducial' runs show the same trends, and Table~\ref{tab:sims} gives data on the entire simulation suite. The left panel shows a comparison to observations, whereas the right panel focuses on comparisons with existing simulation suites. In both cases we also show the results from abundance matching (AM) as implemented by \citet{Read2017}, who used the `field' galaxy stellar mass function from the Sloan Digital Sky Survey (SDSS) and the halo mass function from the cold dark matter Bolshoi simulation \citep[see also][]{Behroozi2013}. Below $M_{200}\sim 5\times 10^9\Msol$ ($M_\star\sim 10^7\Msol$), the relation assumes a power-law extrapolation of the SDSS stellar mass function, as indicated by the dashed lines. 

Note that the simulated stellar masses are taken `as is', i.e. we do not account for any uncertainties from observational colour fitting procedures. Such uncertainties could lead to observational underestimates of a factor up to $\sim 2$ \citep[e.g.][]{Munshi2013}.  Moreover, the simulated $M_{200}$ is the mass of dark matter - we do not include the baryons which at $z=0$ are negligible.

The left panel data compilation consists of local group dwarf galaxies with estimated dynamical masses \citep[][]{Read2017, Read2018}, including galaxies thought to form in halo masses compatible with our simulated galaxies ($M_{200}\sim 10^9\Msol$); LeoT,  Eridanus-II and Carina. As discussed by \citet{Read2018}, the gas rich, star-forming dwarf irregulars have systematically higher estimated galaxy formation efficiencies ($M_\star/M_{200}$) compared to the (non star-forming) satellites. How some low-mass galaxies manage to be star-forming at $z=0$ is not yet clear; see \citet{Wright2018} for a possible line of explanation. In any case, given that our simulated galaxies are quenched reionisation fossils, a direct comparison to the quenched satellites is more appropriate.

The form of the $M_{200}-M_\star$ relation in this low mass regime is highly uncertain, but one can immediately rule out the `No feedback' case as overforming stars by at least one dex. We find that the suppression of stellar mass from supernova feedback (`Hires', and also in the `Fiducial' case which is not plotted) brings galaxies close to AM predictions and local group dwarfs, with $M_\star/M_{200}\sim 2.5 \times 10^{-4}$). Almost an additional order of magnitude suppression comes from radiative feedback (`Hires + RT'), leading to $M_\star/M_{200}\sim 4 \times 10^{-5}$), below the AM extrapolation, but within the uncertainties on Eridanus-II which is the smallest UFD in our observational sample. The model with weak stellar feedback predicts an efficiency of $M_\star/M_{200}\sim 2 \times 10^{-3}$ which is in broad agreement with AM and dIrrs --- but in tension with quenched UFDs, which are the most appropriate point of comparison. 

More observational data on this relationship and its scatter would be helpful to make clearer comparisons to simulations. From the theoretical side, a comparison to existing simulations in the literature is shown in the right panel of Figure~\ref{fig:mstarmhalo}, highlighting that different studies currently make very different predictions for the light-to-mass ratio in the faintest objects. We will discuss this point further in Section~\ref{sect:mstarmhalodisc}, but it should already be clear from the Figure that there is little agreement on the overall effects of feedback in this regime. 

In summary, star formation rates in low mass dwarf galaxies are extremely sensitive to the detailed feedback physics, with radiative feedback contributing significantly to suppression of star formation. Next we demonstrate how these differences arise by studying the density and thermal structure of the ISM.

\begin{figure*}
	\begin{center}
		\includegraphics[width=0.98\textwidth]{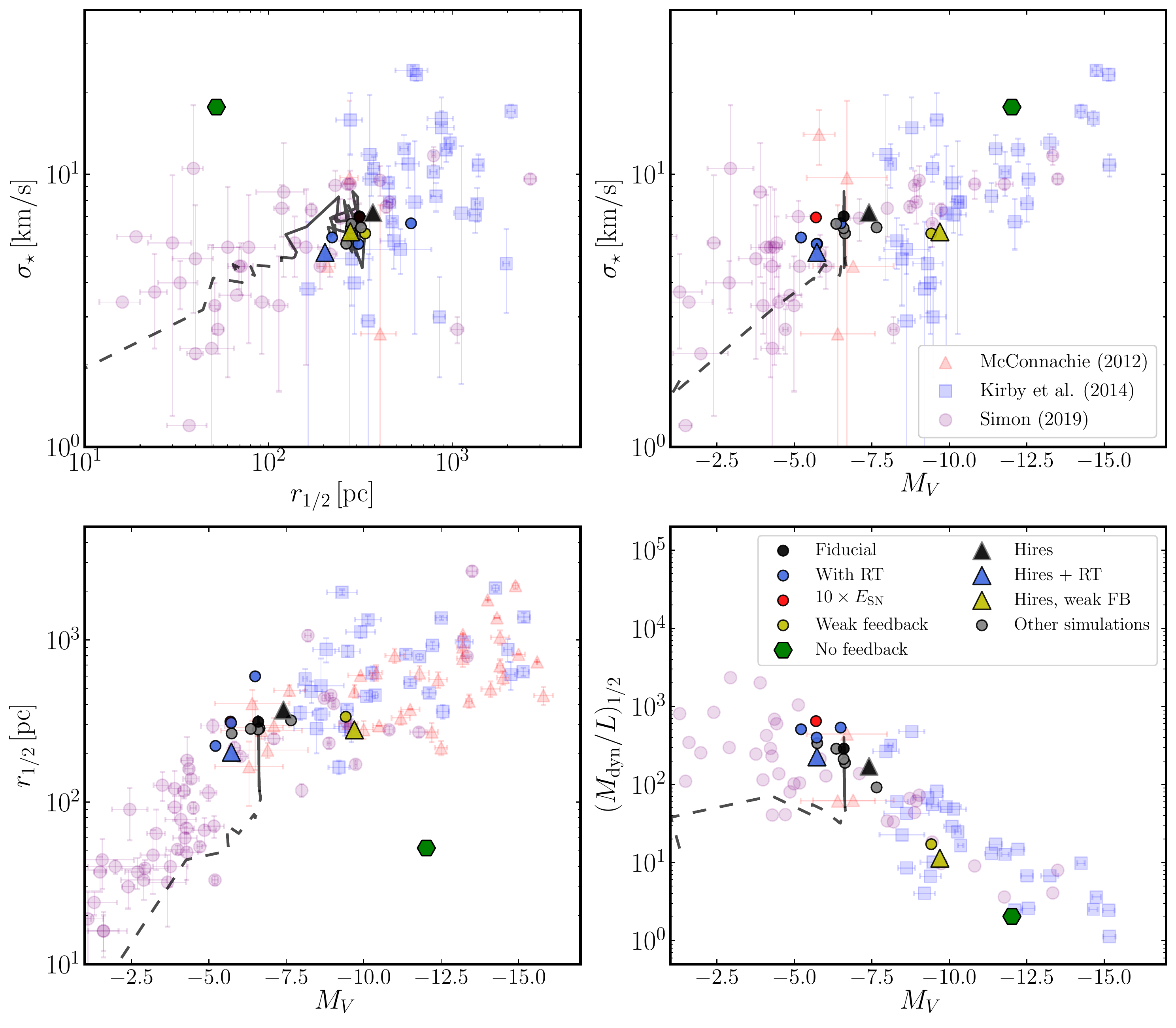} 
		\caption{Structural properties of the simulated dwarfs (bold symbols) and the evolutionary track of the `Fiducial' model (with the dashed portion covering the redshift range $4<z<13$ prior to quenching). Observational data is taken from \citet{McConnachie2012} (red triangles), \citet{Kirby2014} (blue squares) and \citet{Simon2019} (ultra-faints, purple circles). The McConnachie and Kirby et al. data both feature MW, M31 and isolated dwarf galaxies, with some overlap. All simulated galaxies, with the exception of the model without feedback, are found to match observations, highlighting that these scaling relations cannot discriminate strongly between different types of galaxy formation physics.
		}
		\label{fig:scalings}
	\end{center}
\end{figure*}

\subsection{Structural differences of the interstellar medium}
\label{sect:rho}

In Figure~\ref{fig:gasevol}, we show, for a subset of our simulations in the fiducial suite, the mass weighted density (left) and the temperature probability distribution functions (PDFs; right) of the gas in the inner parts of the dark matter halo ($r<0.25 R_{200}$). As the ISM is highly dynamical in low mass galaxies, a single snapshot in time cannot represent the average state of the galaxies. To alleviate this issue\footnote{We note that a more robust analysis, not affected by limited time resolution, would have required mass-weighted PDFs to be computed on-the-fly at \emph{every} simulation time step.}, the PDFs are mass-weighted averages of all simulation outputs in the redshift range $4<z<13$.

Without feedback (grey PDFs), the majority of the ISM's mass is locked up in a population of dense and cold ($T\approx$10~K) star forming clouds reaching densities as high as $n \sim 10^4~{\rm cm}^{-3}$. Enabling SN feedback (black PDF) suppresses the fraction of mass in this phase, and reduces the maximum densities by a factor of 100. Note that most gas, by mass, is still retained in the cold gas phase at $T< 100~$K (below our adopted star formation temperature threshold).

Adding radiative feedback (blue PDFs) has a dramatic effect on the temperature structure, with the peak of the temperature PDF shifted from $T\approx 10~$K to $T\approx 10^4~$K. The large effect of RT on the thermodynamical state of the ISM, and resulting change in the gas mass available for star formation is the primary mechanism for suppressing galactic star formation rates in these models. The suppression is achieved in a relatively `gentle' fashion, without needing to blow the entire interstellar medium from the galaxy. In fact, we find that before reionisation the total gas mass within the galaxy in `Fiducial + RT' is almost identical to the `No feedback' case, despite the two simulations being at opposite extremes in terms of their star formation rates. The total mass in baryons (gas and stars) in the inner halo ($r<0.25 R_{200}$) differ however, with `No feedback' having over 3 times as much baryonic mass at $z\sim 6$ compared to `Fiducial + RT', and almost 10 times the baryonic mass of `Fiducial'.

The gentle suppression of star formation in the `Fiducial + RT' simulation is also evident from the gas density PDF, as it features more dense gas than the other models, except for `Fiducial, no feedback'. Thus star formation is being suppressed without destroying dense clumps. As discussed in Section \ref{sect:intro}, radiative feedback operates immediately upon the birth of massive stars, which enables the remaining dense cloud to be gently heated. By way of contrast, the first SN in a stellar population only explodes after $\sim 4$ Myr, which corresponds to several free-fall times in the dense star forming gas\footnote{Note that stellar winds operate before the first SN explosions in all models, but the low metallicity makes them inefficient at regulating star formation.}. These very different regulation modes will almost certainly have implications for the predicted dark matter distributions in UFDs which we will study in future work; see also Section \ref{sect:discussion}.

Having established the sensitivity of different sub-grid models and numerical resolution on the star formation rates for this isolated dwarf, in the next section we confront our suite of simulations with observations of nearby dwarf galaxies at $z=0$.

\subsection{Dwarf galaxy scaling relations}
\label{sect:scalings}
In Figure~\ref{fig:scalings}, we compare simulated and observed dwarf galaxy properties. We focus on $z=0$ relations between $V$-band magnitudes ($M_V$), half-mass radii ($r_{1/2}$), stellar velocity dispersions ($\sigma_\star$) and dynamical mass-to-light ratios ($M_{\rm dyn}/L$). The  results use the same point styles for the four simulations already shown in Figure~\ref{fig:mstarmhalo}; the remaining simulations are shown as small circles. Simulations with RT switched on are highlighted in blue; the `Fiducial weak feedback' simulation is highlighted in yellow; the `Fiducial $10\times E_{\mathrm{SN}}$' is shown in red; the `Fiducial' simulation is shown in black; and all other simulations in the suite are shown in grey. All results are also reported in Table \ref{tab:sims}. Observational data are compiled from \citet{McConnachie2012}, \citet{Kirby2013}, \citet{Kirby2014},  and the recent compilation for UFDs by \citet{Simon2019}\footnote{\report{These references contain overlapping sources, sometimes with slightly discrepant values for the quantities under consideration. When data is found to overlap, the latest compilation in \citet{Simon2019} take precedence, and for additional sources the data in \citet{Kirby2013} and \citet{Kirby2014} supersede the data in \citet{McConnachie2012}.}}

$V$-band magnitudes are computed using all stars in the main dwarf galaxy halo using the {\small SUNSET}\footnote{Publicly available as a part of the {\small RAMSES} distribution.} ray-tracing code, assuming a Kroupa IMF \citep{Kroupa01}. Velocity dispersions are 1D equivalents, i.e. $\sigma_\star= \sigma_{\rm \star, 3D}/\sqrt{3}$, where $\sigma_{\rm \star, 3D}=\sqrt{\sigma_{\rm \star, x}^2+\sigma_{\rm \star, y}^2+\sigma_{\rm \star, z}^2}$. To allow for closer comparison of the dynamical mass-to-light ratios to the observational data, we follow \citet{Kirby2014} and estimate the dynamical mass using the simple mass estimator:
\begin{equation}
M_{\rm dyn}=3\sigma^2r_{1/2}/G.
\label{eq:MLratio}
\end{equation}

As can be seen in Figure~\ref{fig:scalings}, all simulations, with exception of the outlier `Fiducial, no feedback', have global galaxy properties compatible with observations. In this set of simulations, $M_V$ lies predominantly in the range -5.5 to -6.5, with the simulations including RT being the least luminous. Models with `weak feedback' reach $M_V\sim -9.5$. Despite the range in $M_V$, $z=0$ galaxy sizes and velocity dispersions are found to be quite insensitive to the different galaxy formation scenarios, with $r_{1/2}\sim$ a few 100\,pc and $\sigma_\star\sim5.5-6.5~\kms$ for all galaxies. This weak dependence on final galaxy masses reflects the dominant contribution of the dark matter halo to the gravitational potential. Indeed, dynamical mass-to-light ratios for all galaxies are $10< M_{\rm dyn}/L < 1000$ and are all compatible with observed relations despite the large spread. Again, the model without feedback is an exception to the above, being very compact ($r_{1/2}=51$ pc), with a high velocity dispersion ($\sigma_\star=17.6~\kms$) and self-gravitating ($M_{\rm dyn}/L=2.0$). 

We conclude that it is difficult to rule out -- based solely on the observed scaling relations of dwarf galaxies -- any of our models in which, at minimum, supernova feedback is included. This underlines that, while a large number of studies in the literature have been able to match such relations, they are relatively weak discriminators of the underlying physics governing the faintest dwarfs.

Given the close match to observations, it is interesting to understand whether galaxies evolve `along' observed scaling relations, or if their evolutionary paths are more complex. To this end, Figure~\ref{fig:scalings}  shows evolutionary tracks for the quantities in the representative `Fiducial' model. Dashed lines show $z=13-4$ (i.e. until star formation stops) and solid lines trace the evolution until $z=0$.  The stellar velocity dispersion shows little evolution, varying by at most a factor of 3 over cosmic time. In contrast, the half mass radius evolves significantly, from $r_{1/2}<10\pc$ at $z>10$ to an over one order of magnitude increase in size at the current epoch. After $z=4$, most gas is completely removed due to the background UV radiation, and the galaxy expands by factor of 3, from $r_{1/2}\sim 100$\,pc to $r_{1/2}\sim 300$\,pc. Overall, global quantities are found to more or less evolve along the observed $z=0$ scaling relations, with the possible exception of $r_{1/2}$ and the the early evolution ($z>10$) of the estimated mass-to-light ratio.

\begin{figure*}
	\begin{center}
		\includegraphics[width=0.98\textwidth]{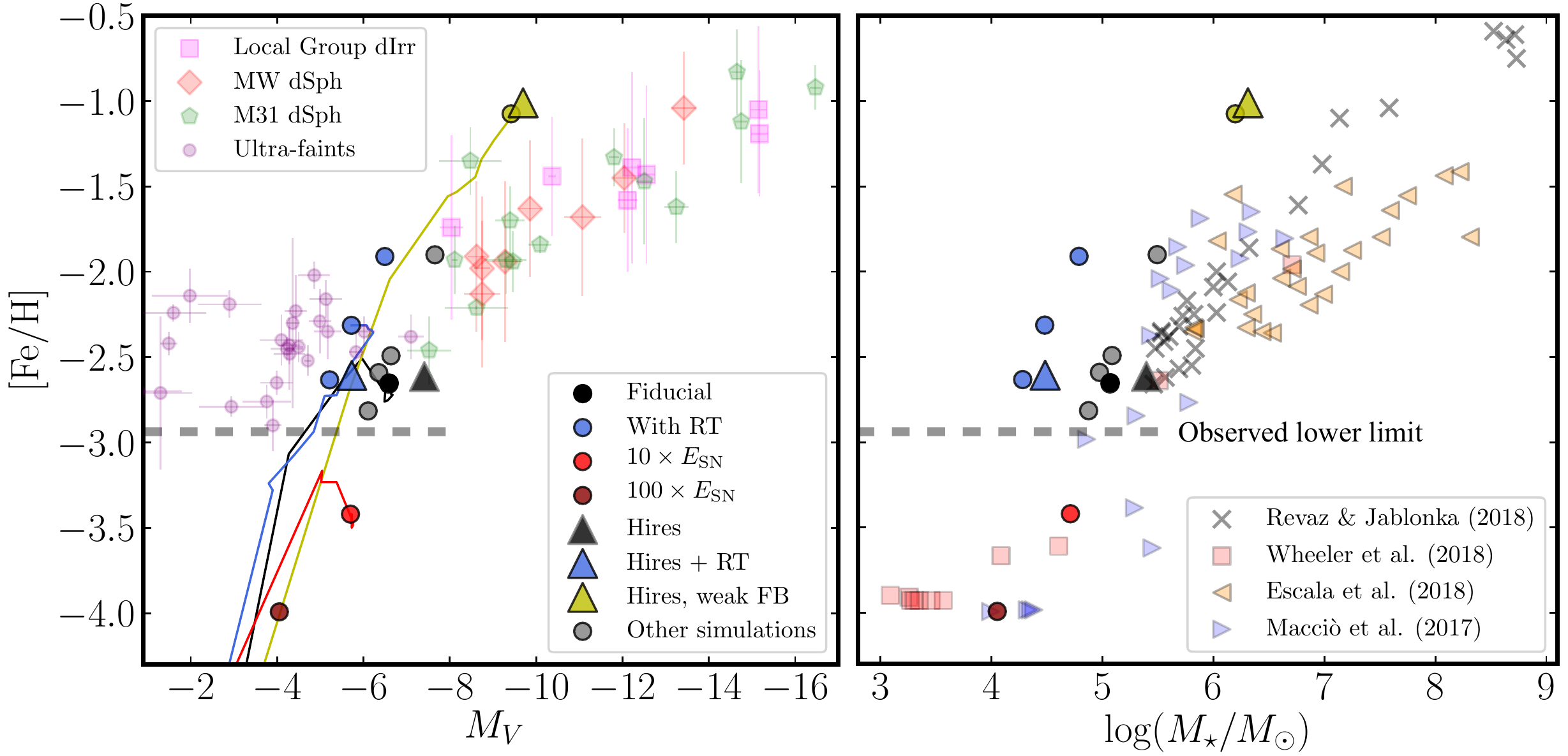} 
		\caption{(Left) [Fe/H] vs. $V$-band magnitude for the simulated dwarfs (bold symbols), compared with observational data taken from \citet{Kirby2013} (classical local group dwarf spheroidal and irregulars: diamonds, pentagons and squares) and \citet{Simon2019} (ultra-faints: circles), with the errorbars showing uncertainties in the mean [Fe/H]. The gray dashed horisontal line indicates the observed lower limit in mean [Fe/H], with the UFD Tucana-II having [Fe/H]=$-2.90^{+0.15}_{-0.16}$ \citep[][]{Chiti2018}, the lowest metallicity in our sample. Each of our simulations is plotted at $z=0$, and the evolution of selected simulations is shown as a line. Models with outflows that are either too strong or too weak fail to match observations, showing that the mass-metallicity relation is a sensitive probe of the feedback mode. (Right) The same metallicities plotted against stellar mass and compared to recent work on zoom simulations of dwarf galaxies in the literature \citep[][]{Maccio2017,Escala2018,Revaz2018}, highlighting that many feedback models struggle to reproduce the observed `plateau' of metal abundances in the faintest dwarfs.
		}
		\label{fig:MZ}
	\end{center}
\end{figure*}

\subsubsection{The stellar mass-metallicity relation}
\label{sect:MZstar}
The final scaling relation we turn to is the stellar mass (or magnitude) vs. metallicity relation (MZR). We henceforth refer to the iron to hydrogen abundance ratio ([Fe/H]) as 'metallicity'.  Figure~\ref{fig:MZ} shows the mean stellar [Fe/H] as a function of $M_V$ (left panel) and $M_\star$ (right panel) for all simulations, using the same point styles as in Figure~\ref{fig:scalings}. In the simulations, chemical abundances are calculated for each star particle following \citep[see also][]{Escala2018},
\begin{equation}
{\rm [Y/X]}=\log_{10}\left(\frac{f_Y/m_Y}{f_X/m_X}\right)-(\log_{10}\epsilon_{Y,\odot}-\log_{10}\epsilon_{X,\odot})
\end{equation}
where $Y$ and $X$ are chemical species, $m_Y$ and $m_X$ are their respective atomic masses, and $f_X$ and $f_Y$ are their respective metal mass fractions. Abundances relative to solar ($\epsilon_{X,\odot}$ and $\epsilon_{Y,\odot}$) are taken from \citet{Asplund2009} (see their table 1). The mean galactic stellar metallicity is computed, motivated by observational measurements of local group dwarf galaxies \citep[e.g.][]{Kirby2013}, according to
\begin{equation}
{\rm [Fe/H]}=\frac{\sum_i^N {\rm [Fe/H]}_i m_{\star,i}}{\sum_i^N m_{\star,i}}
\end{equation}
where [Fe/H]$_i$ is the metallicity of a star particle, $m_{\star, i}$ is the mass of a star particle and $N$ is the number of star particles in the galaxy. 

The left panel contrasts our simulations with data for MW and M31 dwarf spheroidals and local group dwarf irregulars taken from \citet{Kirby2013}, and UFDs from the recent review by \citet{Simon2019}\footnote{With UFDs defined, following the approximate magnitude separation by \citet{Simon2019} (see his figure 5), as galaxies with $M_V>-7.7$.}. Observations suggest a metallicity plateau around [Fe/H]$\sim -2.5$, and we indicate the lower limit observed in mean [Fe/H] (Tucana-II, [Fe/H]=$-2.90^{+0.15}_{-0.16}$, \citealt{Chiti2018}) with a gray dashed line in both panels. This may pose a significant challenge to current numerical simulations which tend to predict near-primordial abundances for the objects with stellar masses significantly below $10^5\,\mathrm{M}_{\odot}$ (right panel), or may indicate that many galaxies at these low luminositites -- that are all satellites of the Milky Way and/or M31 -- are tidally stripped remnants of once larger systems. We will return to this idea in future work.

By way of contrast, some of our feedback setups produce metallicities that are well-matched to observations of low mass dwarfs and UFDs, with [Fe/H]$\sim-2.7$ to $-2$ and $M_V\sim -6.5$ to $-5.5$. The MZR is revealed to be a highly sensitive probe of the physics related to the star formation -- outflow cycle. For example, boosting the amount of energy injected by SNe leads to strong suppression of [Fe/H]. Likewise, weaker stellar feedback models lead to an overestimation of [Fe/H]. This can be understood in terms of our discussion in Section \ref{sect:rho}: {\it explosive} suppression of star formation using supernovae leads to expulsion of enriched gas from the interstellar medium, whereas {\it continuous} suppression of star formation, using mechanisms like UV heating, enables metals to accumulate within the interstellar medium.  

It is also interesting to study the evolutionary tracks in Figure~\ref{fig:MZ}. No version of our galaxy evolves `along' the observed relation, so that star formation quenching at any instance during these formation histories changes the match between theory and observations. Highly explosive suppression of star formation, such as that exhibited by `Fiducial, $10\times E_{\rm SN}$' and  `Fiducial, $100\times E_{\rm SN}$', results in average metallicities\footnote{Allowing for our Pop III metal floor to have solar abundance composition would push up these values to [Fe/H]$\sim -3$ \citep[higher than canonical Pop III to Pop II star formation metallicity thresholds, $Z\sim 10^{-4}~Z_\odot$,][]{Jaacks2018}, which is closer to, but still in tension with, observations. We note that tests (not shown) with pop III floors at $Z\leq 10^{-4}~Z_\odot$ yield close to identical results to those presented here. This means the floor itself, at least in this simulation setup, can be introduced \emph{a posteriori} to understand features such as the observed plateau in [Fe/H].} as low as [Fe/H]$\sim -4$. The opposite is true for the models with `weak feedback', which rapidly evolve to almost one dex above the observed relation.

The power of the observed MZR for distinguishing the feedback models in our simulations is encouraging, since unlike the scaling relations shown in Figure~\ref{fig:scalings}, it may be possible to apply it as a powerful observational diagnostic of galaxy formation physics in the smallest galaxies. In future work (Orkney et al. in prep.), we will explore this using a larger simulation suite. While SN yields \citep[here taken from][]{woosleyweaver1995}, as well as SNIa rates, will introduce uncertainties of factors $\sim 2$ at the low metallicities relevant for UFDs \citep[][]{Wiersma2009}, these are likely sub-dominant compared to the differences between feedback implementations that we find here. A full study will also need to take into account the proximity bias of observed dwarf galaxies with sufficient spectra to derive accurate metallicities \citep{Kirby2013} and environmental effects such as quenching due to infall onto a larger host galaxy. We discuss this further in Section \ref{sect:discussion}. \report{Finally, we note that the MZR in the faintest regime $(M_V\lesssim-3)$ is to date sampled by only a handful of galaxies, and in some cases mean metallicities are poorly determined (for example, Willman 1 has measured metallicities for only 2 stars). Future deep photometric observations and associated spectroscopic followups will determine the degree to which the MZR is an arbiter of galaxy formation physics.}

\section{Discussion}
\label{sect:discussion}
\subsection{Comparison with previous work in the literature}
In this section, we discuss how our results compare to recent dwarf galaxy formation results in the literature, focusing on dwarf galaxy scaling relations and the $M_\star-M_{200}$ relation. Only cosmological simulations reaching a spatial resolution of tens of parsecs are able to resolve the half-mass radii relevant for galaxies forming in haloes of mass $M_{200}\sim 10^8-10^{10}\Msol$ ($r_{1/2}\sim$ few 100\,pc). As a result, we limit our discussion to zoom simulations \citep[although see][for high resolution simulations of a uniform volume to $z=6$]{Rosdahl2018}.

\subsubsection{Effects of feedback on simulated dwarf galaxies}
Several authors, using a variety of numerical methods, star formation and feedback prescriptions, report on global properties of simulated dwarfs forming in $M_{200}\sim 10^9 \Msol$ haloes \citep[including][]{Wheeler2015, Onorbe2015,Fitts2017,Maccio2017,Munshi2017,Revaz2018}. The general picture is that modern simulations produce galaxies compatible with observed scaling relations, with $r_{1/2}\sim 0.1-1~\kpc$, $\sigma_{\star}\sim 5-10~\kms$ and stellar masses $M_\star\sim 10^4-10^6\Msol$. All successful models include subgrid models for stellar feedback that are strong enough to regulate star formation. This keeps dynamical mass-to-light ratios high, allowing stellar velocity dispersions and half-mass radii to be set by the dark matter halo alone.

While most simulations agree on this basic result, \citet{SmithSijacki2019} give an important caveat. Despite the  inclusion of a momentum-capturing SN model similar to ours (see Section \ref{sect:galformphysics}), their simulations all suffer from over-cooling, leading to dwarf galaxy properties similar to our `Fiducial, No Feedback' case. This issue was only mitigated when stars in the \citet{SmithSijacki2019} simulations were modelled to form at a high star formation efficiency per free-fall time, $\epsilon_{\rm ff}=100\%$, leading to a more clustered injection of stellar feedback. Nonetheless, while dynamical scaling relations do capture some important aspects of the interaction between feedback algorithms (e.g. `blastwave' vs. momentum based feedback) and the clustering of star formation (see e.g. the discussion in \citealt{AgertzKravtsov2015}), we find that metal enrichment is a considerably more sensitive probe of the different feedback models that we have studied in this work.

\citet{Wheeler2015} and \citet{Onorbe2015} used the FIRE model of feedback \citep[][see also the updated FIRE2 model, \citealt{Hopkins2018}]{Hopkins2014}, which treats a range of feedback processes, similar to our model, including a subgrid model of radiative feedback that includes both radiation pressure and ionising radiation. The Smoothed-Particle-Hydrodynamics (SPH) simulations by \citet{Maccio2017} and \citet{Revaz2018} both adopted the `blastwave' feedback model \citep[][]{Stinson06} that prohibits gas cooling around SNe explosions for an extended period of time, in order to allow for efficient coupling to the ISM. In addition, \citet{Maccio2017} made use of a phenomenological model  for the effect of UV radiation, `Early Stellar Feedback'\footnote{In this, 10\% of the UV luminosity of a stellar population -- typically $10^{52}$ ergs (10 times the canonical SN explosion energy) -- is injected as thermal energy before any SN events take place \citep{Stinson2013}.}, which is likely the reason for them finding lower stellar masses than \citet{Revaz2018} ($M_\star \sim 10^4\Msol$ compared to $M_\star\sim 10^5\Msol - 10^6\Msol$ for galaxies of mass $M_{200}=10^9\Msol$). The right panel of Figure~\ref{fig:MZ} shows that the [Fe/H]-$M_\star$ relation in \citet{Maccio2017} diverge from the observed one \citep[][]{Kirby2014} below $M_\star\sim10^6\Msol$. 

Similar results were recently found by \citet{Wheeler2018} also shown in the right panel of Figure~\ref{fig:MZ}, at a similar numerical resolution to our `Hires' simulations; for $M_\star\lesssim 10^5~\Msol$, their galaxies have mean [Fe/H]$<-3.5$, with UFDs $M_\star\sim 10^3-10^4~\Msol$ never enriching at all. This possibly indicates that either star formation is shut down too early in their simulated dwarfs due to overly efficient feedback, or that the IGM around star forming UFDs, at least in the local group, was significantly more pre-enriched by Pop III stars than traditionally thought, as well as predicted by galaxy simulations with Pop III enrichment (e.g. \citealt{Vandenbrouke2016}, but see \citealt{Jaacks2018}), bringing their simulated UFDs closer to the observed [Fe/H] lower limit found for faint satellite galaxies.

\subsubsection{The  $M_\star-M_{200}$ relation}
\label{sect:mstarmhalodisc}
At present, there is no clear theoretical consensus on expectations for $M_\star$ for halo masses $M_{200}$ below $\simeq 10^{10}\,\Msol$. While individual simulation efforts with fixed star formation and feedback prescriptions tend to produce well-defined $M_\star$ -- $M_{200}$ relations \citep[e.g.][]{Wheeler2015,Fitts2017,Munshi2017, Wheeler2018}, there are major differences between groups \citep[$\sim 2$ dex, see e.g.][]{Garrison-Kimmel2017}. 

 The right panel of Figure~\ref{fig:mstarmhalo} collects some examples from the literature. Work using the FIRE feedback and star formation models predicts steeply decreasing galaxy formation efficiencies below $M_{200}\sim 10^{10}\Msol$, with $M_\star/M_{200}< 10^{-5}$ around $M_{200}\sim10^{9}\Msol$ \citep[close to a few $\times10^{-6}$, also found in the more recent work by][]{Wheeler2018}. As such, the FIRE simulations predict the lowest galaxy formation efficiencies\footnote{We note that a few of the \citet{Wetzel2016} `Latte' dwarfs are in line with our results} considered here \citep[together with][]{Maccio2017}, possibly due to strong feedback and the inclusion of ISM heating sources such as photo-electric heating (see Section \ref{sect:limitations}). Our `Hires+RT' model predicts $M_\star/M_{200}\sim 5\times 10^{-5}$, one order of magnitude higher.
 
 The data from the cosmological zoom SPH simulations in \citet{Munshi2017} show a variety of efficiencies, with results in general agreement with our entire simulation suite. By extending their work to include molecular hydrogen based star formation, akin to our approach, \citet{Munshi2018} found that stars can only form at densities high enough to allow for gas self-shielding ($n\sim 1000~m_{\rm H}~{\rm cm}^{-3}$), in agreement with our findings for the `Fiducial + RT, H$_2$ based SF' simulation. This led to a large suppression of the number of galaxies formed below $M_{200}\sim 10^{9}\Msol$. We found essentially no effect when adopting a H$_2$-based prescription compared to our fiducial approach, which likely stems from us anyway restricting star formation to dense ($n>300~{\rm cm}^{-3}$) and cold gas ($T<100$ K) that roughly captures the environments where molecular hydrogen formation is possible. Note that we have only studied the effect of H$_2$ physics for one particular dwarf assembly history, and a larger suite may reveal if H$_2$ based star formation has an impact. Furthermore, the insensitivity can also stem from us already employing a high density threshold for star formation in all simulations, or that we are not modelling low enough halo masses, which we leave for a future investigation. When coupled with RT, the impact of H$_2$-based star formation led to a $\sim 50\%$ difference in the $z=0$ stellar masses.
 
There is likely to be a physical spread in $M_\star$ at a given $M_{200}$ resulting from the diversity of mass growth histories and the different environments experienced by dwarfs \citep[e.g.][]{Garrison-Kimmel2017,Fitts2017,Munshi2017,Wright2018}. Satellite dwarfs experience dark matter and stellar mass loss due to tidal stripping, biasing them to higher $M_\star/M_{200}$ ratios \citep[][]{Sawala2015}. However, it should be emphasised that the current scatter in results from the numerical literature is almost certainly dominated by pure feedback discrepancies rather than any systematic differences in the objects being simulated. 

In summary, the diversity in $M_\star-M_{200}$ at `the edge of galaxy formation' found in the literature reflects \report{in part} specific choices in terms of numerical methods, including flavours of subgrid feedback and star formation prescriptions, as demonstrated by the large scatter found in this paper for a \emph{single} choice of initial condition. In order to interpret observations robustly, both the feedback uncertainty and history variations \citep[see also][]{Revaz2018} will have to be taken into account; we investigate this in a companion study \citep[][]{Rey2019}.

\subsection{Simulation limitations}
\label{sect:limitations}
Having demonstrated that radiative feedback plays a crucial role in regulating the rate of star formation in ultra-faint dwarfs, we now turn to processes we have omitted. Feedback models are still far from complete, but a census of the leading-order missing ingredients and their likely effects helps to interpret the current state-of-the-art.

First, we do not model the effect of photoelectric heating (PEH), i.e. the heating by far-UV photons from young stars as they liberate electrons from dust grains \citep[][]{Draine1978}. Although this effect is sub-dominant to UV heating, \citet{Forbes2016} carried out non-cosmological simulations of dwarf galaxies and found PEH to significantly reduce dwarf galaxy stellar masses, more so than supernovae feedback. On the other hand, \citet{Hu2017} suggested that the results of \citet{Forbes2016} arose due to incorrect cooling rates in the self-shielded gas. It therefore seems likely that the non-linear coupling between SNe, RT, and dispersal of dense gas, renders PEH subdominant to SN feedback. \report{Furthermore, as the dust-to-gas ratio decreases with metallicity \citep[super-linearly below 0.1$Z_\odot$,][]{Fisher2014}, PEH will most likely be a sub-dominant effect in the low metallicity environments studied in this work.} Nonetheless, further work in this area is warranted. 

Second, we do not include the effect of resonantly scattered Lyman-$\alpha$ photons. Self-consistently modelling the momentum transfer from the scattering of Ly-$\alpha$ photons onto the gas is computationally challenging. \citet{Kimm2018} implemented a subgrid model of this effect in {\small RAMSES-RT}, and argued that in metal poor systems, Ly-$\alpha$ photons impart momentum comparable to SNe. In their simulations of an idealised disc embedded in $M_{200}=10^{10}\Msol$ halo, with supernova and RT physics similar to that adopted in this work, star formation rates were reduced by a factor of two. This effect has not yet been studied in a cosmological context. 
Another missing feedback effect is cosmic rays (CRs). Work by \citet{Booth2013} \citep[see also][]{Hanasz2013,SalemBryan2014,Chan2018} demonstrated that the pressure gradients generated by cosmic rays can lead to winds with high mass loading factors, on the order of $\sim 10$, in dwarf galaxies. A full treatment CR driven winds requires magnetohydrodynamics as well as anisotropic cosmic ray diffusion coefficients. Furthermore, the impact of CRs on galactic wind properties depend strongly on the value of the diffusion coefficients, which are only empirically constrained withinn a factor of 10 for the Milky Way's ISM, making conclusions on the impact of CRs less robust, at least currentely, than e.g. supernova feedback.

Finally, each star particle formed in our simulations is assumed to be an SSP with a \citet{Kroupa01} IMF. In ultra-faint galaxies, the way in which the IMF is populated likely matters, as stochasticity can affect the local level of gas heating from SNe and hence burstiness of star formation \citep[see recent work by][]{Applebaum2018}. Although highly uncertain, if the IMF depends on the local ISM environment by e.g. becoming top-heavy at the low metallicities relevant for UFDs \citep[as suggested by][]{Geha2013}, this can affect galaxy formation in non-trivial ways (Prgomet et al. in prep.). Furthermore, binary star physics can provide additional sources of feedback that have not been modelled here \citep[e.g.][]{Jeon2015}.

We leave the investigation of how the above processes impact galaxy formation in low mass haloes for future work.

\section{Conclusions}
\label{sect:conclusions}
We have carried out cosmological zoom simulations with coupled radiation and hydrodynamics to study the formation of ultra-faint dwarf (UFD) galaxies. We studied galaxy formation in a dark matter halos, forming in relative isolation, with $z=0$ virial masses $M_{200}=10^9\Msol$, such that the cosmic UV background quenched star formation by $z\sim 4$. The simulations reached a mass and spatial resolution of $\sim 20\Msol$ and $\sim 3$\,parsecs. Using a single realization of the Gaussian random initial conditions for our dwarf, we investigated the sensitivity of observed galaxy properties to the adopted supernova feedback model, UV background, molecular hydrogen (H$_2$) physics, numerical resolution and multifrequency radiative transfer (RT). Our key findings are as follows:

\begin{itemize}
\item Supernova (SN) feedback lowers galaxy masses by two orders of magnitude, from $M_\star\sim 10^7\Msol$ when feedback is not included, to  $M_\star\sim10^5\Msol$. Radiative feedback, here self-consistently modeled using multi-frequency radiative transfer (RT), lowers the galaxy formation efficiency by almost an additional order of magnitude, bringing stellar masses closer to a ${\rm few}\times 10^4\Msol$, similar to local group ultra-faint dwarfs such as Eridanus-II. 

\item In models without radiative feedback, we find that dwarf galaxy formation is regulated by vigorous outflows, with much of the interstellar medium being dispersed during starburst episodes. This picture changes when RT is introduced. Radiative feedback acts to keep more gas in a warm ($\sim 10^4$ K) non-star forming state, leading to less gas accretion and collapse, and, as a result, less vigorous galactic scale outflows, at all times.

\item All of our simulations with efficient feedback are in agreement with dynamical measurements of $M_\star$ vs. $M_{200}$ for local group dwarf satellites and UFDs, with $M_\star/M_{200}\sim 10^{-4}$. Molecular hydrogen based star formation and changes to the UV background modifies final stellar masses by at most a factor of two. 

\item All simulations, even those with artificially weak feedback and $M_\star/M_{200}>10^{-3}$, lie within the scatter of observed scaling relations for $V$-band magnitudes, half mass radii, stellar velocity dispersions and dynamical mass-to-light ratios from local group dwarf irregulars, spheroidals and UFDs, with $r_{1/2}\sim 200-400\pc$ and $\sigma_{\star}\sim 5-7~\kms$. We find that this insensitivity arises due to structural scaling relations predominantly being set by the host dark matter halo in galaxies with high mass-to-light-ratios.

\item We find that the stellar mass-metallicity relation, \report{based on currently available data}, can differentiate galaxy formation models, as simulations fail to match observations whenever SN feedback is artificially strong or weak. This highlights that the success of galaxy formation models depends not necessarily on how many observables they can match, but rather on whether they can match a few key observables -- in this case the stellar mass-metallicity relation -- that are sensitive to changes in the sub-grid physics model.
\end{itemize}

Our work demonstrates that the enrichment of dwarf galaxies is a more powerful discriminant of feedback processes than any of the other observed scaling relations. The tendency for simulations of ultra-faint dwarfs in the literature to be under-enriched may suggest that their feedback models may be excessively violent, unphysically ejecting enriched gas at high redshift, or that the IGM was pre-enriched to mucher higher levels than traditionally thought. In future work \citep[][Orkney et al. in prep]{Rey2019}, we will study a wider range of assembly histories and halo masses in order to understand the generality of this conclusion, as well as to probe how accretion history and environment shapes both the metallicity and dynamics of the ultra-faint dwarf population.

\section*{acknowledgments}
\report{The authors thank the referee Chia-Yu Hu for his constructive comments that improved the quality of the paper. We thank Coral Wheeler, Joshua Simon, Andrew Wetzel, Andrea Macci\`o, Yvez Revaz and Alyson Brooks for helpful comments.}

OA acknowledges support from the Swedish Research Council (grant 2014-5791) and the Knut and Alice Wallenberg Foundation.  AP is supported by the Royal Society. MR acknowledges support from the Perren Fund and the IMPACT fund. JR acknowledges support from the ORAGE project from the Agence Nationale de la Recherche under grant ANR-14-CE33-0016-03.

This work was partially enabled by support from the UCL Cosmoparticle Initiative. This work was performed in part using the DiRAC Data Intensive service at Leicester, operated by the University of Leicester IT Services, which forms part of the STFC DiRAC HPC Facility (www.dirac.ac.uk). The equipment was funded by BEIS capital funding via STFC capital grants ST/K000373/1 and ST/R002363/1 and STFC DiRAC Operations grant ST/R001014/1. DiRAC is part of the National e-Infrastructure. This work also used the COSMA Data Centric system at Durham University, operated by the Institute for Computational Cosmology on behalf of the STFC DiRAC HPC Facility (www.dirac.ac.uk). This equipment was funded by a BIS National E-infrastructure capital grant ST/K00042X/1, DiRAC Operations grant ST/K003267/1 and Durham University. DiRAC is part of the National E-Infrastructure. This work was also supported by a grant from the Swiss National Supercomputing Centre (CSCS) under project ID s890. Finally, a large number of simulations for this work were performed on computational resources at LUNARC, the center for scientific and technical computing at Lund University, thanks to financial support from the Royal Physiographic Society of Lund.

\footnotesize{
\bibliography{ms.bbl}

\begin{thebibliography}{}
\makeatletter
\relax
\def\mn@urlcharsother{\let\do\@makeother \do\$\do\&\do\#\do\^\do\_\do\%\do\~}
\def\mn@doi{\begingroup\mn@urlcharsother \@ifnextchar [ {\mn@doi@}
  {\mn@doi@[]}}
\def\mn@doi@[#1]#2{\def\@tempa{#1}\ifx\@tempa\@empty \href
  {http://dx.doi.org/#2} {doi:#2}\else \href {http://dx.doi.org/#2} {#1}\fi
  \endgroup}
\def\mn@eprint#1#2{\mn@eprint@#1:#2::\@nil}
\def\mn@eprint@arXiv#1{\href {http://arxiv.org/abs/#1} {{\tt arXiv:#1}}}
\def\mn@eprint@dblp#1{\href {http://dblp.uni-trier.de/rec/bibtex/#1.xml}
  {dblp:#1}}
\def\mn@eprint@#1:#2:#3:#4\@nil{\def\@tempa {#1}\def\@tempb {#2}\def\@tempc
  {#3}\ifx \@tempc \@empty \let \@tempc \@tempb \let \@tempb \@tempa \fi \ifx
  \@tempb \@empty \def\@tempb {arXiv}\fi \@ifundefined
  {mn@eprint@\@tempb}{\@tempb:\@tempc}{\expandafter \expandafter \csname
  mn@eprint@\@tempb\endcsname \expandafter{\@tempc}}}

\bibitem[\protect\citeauthoryear{{Abel}, {Anninos}, {Norman}  \&
  {Zhang}}{{Abel} et~al.}{1998}]{1998ApJ...508..518A}
{Abel} T.,  {Anninos} P.,  {Norman} M.~L.,   {Zhang} Y.,  1998, \mn@doi [\apj]
  {10.1086/306410}, \href {http://adsabs.harvard.edu/abs/1998ApJ...508..518A}
  {508, 518}

\bibitem[\protect\citeauthoryear{{Agertz} \& {Kravtsov}}{{Agertz} \&
  {Kravtsov}}{2015}]{AgertzKravtsov2015}
{Agertz} O.,  {Kravtsov} A.~V.,  2015, \mn@doi [\apj]
  {10.1088/0004-637X/804/1/18}, \href
  {http://adsabs.harvard.edu/abs/2015ApJ...804...18A} {804, 18}

\bibitem[\protect\citeauthoryear{{Agertz} \& {Kravtsov}}{{Agertz} \&
  {Kravtsov}}{2016}]{AgertzKravtsov2016}
{Agertz} O.,  {Kravtsov} A.~V.,  2016, \mn@doi [\apj]
  {10.3847/0004-637X/824/2/79}, \href
  {http://adsabs.harvard.edu/abs/2016ApJ...824...79A} {824, 79}

\bibitem[\protect\citeauthoryear{{Agertz}, {Kravtsov}, {Leitner}  \&
  {Gnedin}}{{Agertz} et~al.}{2013}]{Agertz2013}
{Agertz} O.,  {Kravtsov} A.~V.,  {Leitner} S.~N.,   {Gnedin} N.~Y.,  2013,
  \mn@doi [\apj] {10.1088/0004-637X/770/1/25}, \href
  {http://adsabs.harvard.edu/abs/2013ApJ...770...25A} {770, 25}

\bibitem[\protect\citeauthoryear{{Agertz}, {Romeo}  \& {Grisdale}}{{Agertz}
  et~al.}{2015}]{AgertzRomeoGrisdale2015}
{Agertz} O.,  {Romeo} A.~B.,   {Grisdale} K.,  2015, \mn@doi [\mnras]
  {10.1093/mnras/stv440}, \href
  {http://adsabs.harvard.edu/abs/2015MNRAS.449.2156A} {449, 2156}

\bibitem[\protect\citeauthoryear{{Amorisco}}{{Amorisco}}{2017}]{Amorisco2017}
{Amorisco} N.~C.,  2017, \mn@doi [\apj] {10.3847/1538-4357/aa745f}, \href
  {http://adsabs.harvard.edu/abs/2017ApJ...844...64A} {844, 64}

\bibitem[\protect\citeauthoryear{{Applebaum}, {Brooks}, {Quinn}  \&
  {Christensen}}{{Applebaum} et~al.}{2018}]{Applebaum2018}
{Applebaum} E.,  {Brooks} A.~M.,  {Quinn} T.~R.,   {Christensen} C.~R.,  2018,
  preprint, \href {http://adsabs.harvard.edu/abs/2018arXiv181100022A} {}
  (\mn@eprint {arXiv} {1811.00022})

\bibitem[\protect\citeauthoryear{{Asplund}, {Grevesse}, {Sauval}  \&
  {Scott}}{{Asplund} et~al.}{2009}]{Asplund2009}
{Asplund} M.,  {Grevesse} N.,  {Sauval} A.~J.,   {Scott} P.,  2009, \mn@doi
  [\araa] {10.1146/annurev.astro.46.060407.145222}, \href
  {http://adsabs.harvard.edu/abs/2009ARA%26A..47..481A} {47, 481}

\bibitem[\protect\citeauthoryear{{Barkana} \& {Loeb}}{{Barkana} \&
  {Loeb}}{1999}]{1999ApJ...523...54B}
{Barkana} R.,  {Loeb} A.,  1999, \mn@doi [\apj] {10.1086/307724}, \href
  {http://adsabs.harvard.edu/abs/1999ApJ...523...54B} {523, 54}

\bibitem[\protect\citeauthoryear{{Behroozi}, {Wechsler}  \&
  {Conroy}}{{Behroozi} et~al.}{2013}]{Behroozi2013}
{Behroozi} P.~S.,  {Wechsler} R.~H.,   {Conroy} C.,  2013, \mn@doi [\apj]
  {10.1088/0004-637X/770/1/57}, \href
  {http://adsabs.harvard.edu/abs/2013ApJ...770...57B} {770, 57}

\bibitem[\protect\citeauthoryear{{Benincasa}, {Wadsley}, {Couchman}  \&
  {Keller}}{{Benincasa} et~al.}{2016}]{Benincasa2016}
{Benincasa} S.~M.,  {Wadsley} J.,  {Couchman} H.~M.~P.,   {Keller} B.~W.,
  2016, \mn@doi [\mnras] {10.1093/mnras/stw1741}, \href
  {http://adsabs.harvard.edu/abs/2016MNRAS.462.3053B} {462, 3053}

\bibitem[\protect\citeauthoryear{{Benitez-Llambay}, {Frenk}, {Ludlow}  \&
  {Navarro}}{{Benitez-Llambay} et~al.}{2018}]{2018arXiv181004186B}
{Benitez-Llambay} A.,  {Frenk} C.~S.,  {Ludlow} A.~D.,   {Navarro} J.~F.,
  2018, arXiv e-prints, \href
  {http://adsabs.harvard.edu/abs/2018arXiv181004186B} {}

\bibitem[\protect\citeauthoryear{{Benson}, {Frenk}, {Lacey}, {Baugh}  \&
  {Cole}}{{Benson} et~al.}{2002}]{2002MNRAS.333..177B}
{Benson} A.~J.,  {Frenk} C.~S.,  {Lacey} C.~G.,  {Baugh} C.~M.,   {Cole} S.,
  2002, \mn@doi [\mnras] {10.1046/j.1365-8711.2002.05388.x}, \href
  {http://adsabs.harvard.edu/abs/2002MNRAS.333..177B} {333, 177}

\bibitem[\protect\citeauthoryear{{Bigiel}, {Leroy}, {Walter}, {Brinks}, {de
  Blok}, {Madore}  \& {Thornley}}{{Bigiel} et~al.}{2008}]{bigiel2008}
{Bigiel} F.,  {Leroy} A.,  {Walter} F.,  {Brinks} E.,  {de Blok} W.~J.~G.,
  {Madore} B.,   {Thornley} M.~D.,  2008, \mn@doi [\aj]
  {10.1088/0004-6256/136/6/2846}, \href
  {http://adsabs.harvard.edu/abs/2008AJ....136.2846B} {136, 2846}

\bibitem[\protect\citeauthoryear{{Bland-Hawthorn}, {Sutherland}  \&
  {Webster}}{{Bland-Hawthorn} et~al.}{2015}]{2015ApJ...807..154B}
{Bland-Hawthorn} J.,  {Sutherland} R.,   {Webster} D.,  2015, \mn@doi [\apj]
  {10.1088/0004-637X/807/2/154}, \href
  {http://adsabs.harvard.edu/abs/2015ApJ...807..154B} {807, 154}

\bibitem[\protect\citeauthoryear{{Booth}, {Agertz}, {Kravtsov}  \&
  {Gnedin}}{{Booth} et~al.}{2013}]{Booth2013}
{Booth} C.~M.,  {Agertz} O.,  {Kravtsov} A.~V.,   {Gnedin} N.~Y.,  2013,
  \mn@doi [\apjl] {10.1088/2041-8205/777/1/L16}, \href
  {http://adsabs.harvard.edu/abs/2013ApJ...777L..16B} {777, L16}

\bibitem[\protect\citeauthoryear{{Bose} et~al.,}{{Bose}
  et~al.}{2018}]{2018arXiv181003635B}
{Bose} S.,  et~al., 2018, arXiv e-prints, \href
  {http://adsabs.harvard.edu/abs/2018arXiv181003635B} {}

\bibitem[\protect\citeauthoryear{{Bovill} \& {Ricotti}}{{Bovill} \&
  {Ricotti}}{2009}]{2009ApJ...693.1859B}
{Bovill} M.~S.,  {Ricotti} M.,  2009, \mn@doi [\apj]
  {10.1088/0004-637X/693/2/1859}, \href
  {http://adsabs.harvard.edu/abs/2009ApJ...693.1859B} {693, 1859}

\bibitem[\protect\citeauthoryear{{Bovill} \& {Ricotti}}{{Bovill} \&
  {Ricotti}}{2011}]{2011ApJ...741...18B}
{Bovill} M.~S.,  {Ricotti} M.,  2011, \mn@doi [\apj]
  {10.1088/0004-637X/741/1/18}, \href
  {http://adsabs.harvard.edu/abs/2011ApJ...741...18B} {741, 18}

\bibitem[\protect\citeauthoryear{{Brown} et~al.,}{{Brown}
  et~al.}{2014}]{2014ApJ...796...91B}
{Brown} T.~M.,  et~al., 2014, \mn@doi [\apj] {10.1088/0004-637X/796/2/91},
  \href {http://adsabs.harvard.edu/abs/2014ApJ...796...91B} {796, 91}

\bibitem[\protect\citeauthoryear{{Bruzual} \& {Charlot}}{{Bruzual} \&
  {Charlot}}{2003}]{BC03}
{Bruzual} G.,  {Charlot} S.,  2003, \mn@doi [\mnras]
  {10.1046/j.1365-8711.2003.06897.x}, \href
  {https://ui.adsabs.harvard.edu/\#abs/2003MNRAS.344.1000B} {344, 1000}

\bibitem[\protect\citeauthoryear{{Bullock} \& {Boylan-Kolchin}}{{Bullock} \&
  {Boylan-Kolchin}}{2017}]{2017ARA&A..55..343B}
{Bullock} J.~S.,  {Boylan-Kolchin} M.,  2017, \mn@doi [\araa]
  {10.1146/annurev-astro-091916-055313}, \href
  {http://adsabs.harvard.edu/abs/2017ARA%26A..55..343B} {55, 343}

\bibitem[\protect\citeauthoryear{{Bullock}, {Kravtsov}  \&
  {Weinberg}}{{Bullock} et~al.}{2000}]{2000ApJ...539..517B}
{Bullock} J.~S.,  {Kravtsov} A.~V.,   {Weinberg} D.~H.,  2000, \mn@doi [\apj]
  {10.1086/309279}, \href {http://adsabs.harvard.edu/abs/2000ApJ...539..517B}
  {539, 517}

\bibitem[\protect\citeauthoryear{{Chan}, {Keres}, {Hopkins}, {Quataert}, {Su},
  {Hayward}  \& {Faucher-Giguere}}{{Chan} et~al.}{2018}]{Chan2018}
{Chan} T.~K.,  {Keres} D.,  {Hopkins} P.~F.,  {Quataert} E.,  {Su} K.-Y.,
  {Hayward} C.~C.,   {Faucher-Giguere} C.-A.,  2018, arXiv e-prints, \href
  {http://adsabs.harvard.edu/abs/2018arXiv181210496C} {}

\bibitem[\protect\citeauthoryear{{Chiti}, {Frebel}, {Ji}, {Jerjen}, {Kim}  \&
  {Norris}}{{Chiti} et~al.}{2018}]{Chiti2018}
{Chiti} A.,  {Frebel} A.,  {Ji} A.~P.,  {Jerjen} H.,  {Kim} D.,   {Norris}
  J.~E.,  2018, \mn@doi [\apj] {10.3847/1538-4357/aab4fc}, \href
  {https://ui.adsabs.harvard.edu/\#abs/2018ApJ...857...74C} {857, 74}

\bibitem[\protect\citeauthoryear{{Christensen}, {Governato}, {Quinn}, {Brooks},
  {Shen}, {McCleary}, {Fisher}  \& {Wadsley}}{{Christensen}
  et~al.}{2014}]{Christensen2014}
{Christensen} C.~R.,  {Governato} F.,  {Quinn} T.,  {Brooks} A.~M.,  {Shen} S.,
   {McCleary} J.,  {Fisher} D.~B.,   {Wadsley} J.,  2014, \mn@doi [\mnras]
  {10.1093/mnras/stu399}, \href
  {https://ui.adsabs.harvard.edu/\#abs/2014MNRAS.440.2843C} {440, 2843}

\bibitem[\protect\citeauthoryear{{Contenta} et~al.,}{{Contenta}
  et~al.}{2018}]{Contenta2018}
{Contenta} F.,  et~al., 2018, \mn@doi [\mnras] {10.1093/mnras/sty424}, \href
  {http://adsabs.harvard.edu/abs/2018MNRAS.476.3124C} {476, 3124}

\bibitem[\protect\citeauthoryear{{Crain} et~al.,}{{Crain}
  et~al.}{2015}]{2015MNRAS.450.1937C}
{Crain} R.~A.,  et~al., 2015, \mn@doi [\mnras] {10.1093/mnras/stv725}, \href
  {http://adsabs.harvard.edu/abs/2015MNRAS.450.1937C} {450, 1937}

\bibitem[\protect\citeauthoryear{{Dalla Vecchia} \& {Schaye}}{{Dalla Vecchia}
  \& {Schaye}}{2008}]{2008MNRAS.387.1431D}
{Dalla Vecchia} C.,  {Schaye} J.,  2008, \mn@doi [\mnras]
  {10.1111/j.1365-2966.2008.13322.x}, \href
  {http://adsabs.harvard.edu/abs/2008MNRAS.387.1431D} {387, 1431}

\bibitem[\protect\citeauthoryear{{Decataldo}, {Pallottini}, {Ferrara},
  {Vallini}  \& {Gallerani}}{{Decataldo} et~al.}{2019}]{Decataldo2019}
{Decataldo} D.,  {Pallottini} A.,  {Ferrara} A.,  {Vallini} L.,   {Gallerani}
  S.,  2019, \mn@doi [\mnras] {10.1093/mnras/stz1527}, \href
  {https://ui.adsabs.harvard.edu/abs/2019MNRAS.487.3377D} {487, 3377}

\bibitem[\protect\citeauthoryear{{Dekel} \& {Silk}}{{Dekel} \&
  {Silk}}{1986}]{1986ApJ...303...39D}
{Dekel} A.,  {Silk} J.,  1986, \mn@doi [\apj] {10.1086/164050}, \href
  {http://adsabs.harvard.edu/abs/1986ApJ...303...39D} {303, 39}

\bibitem[\protect\citeauthoryear{{Draine}}{{Draine}}{1978}]{Draine1978}
{Draine} B.~T.,  1978, \mn@doi [\apjs] {10.1086/190513}, \href
  {http://adsabs.harvard.edu/abs/1978ApJS...36..595D} {36, 595}

\bibitem[\protect\citeauthoryear{{Dutton}, {Macci{\`o}}, {Buck}, {Dixon},
  {Blank}  \& {Obreja}}{{Dutton} et~al.}{2018}]{2018arXiv181110625D}
{Dutton} A.~A.,  {Macci{\`o}} A.~V.,  {Buck} T.,  {Dixon} K.~L.,  {Blank} M.,
  {Obreja} A.,  2018, arXiv e-prints, \href
  {http://adsabs.harvard.edu/abs/2018arXiv181110625D} {}

\bibitem[\protect\citeauthoryear{{Efstathiou}}{{Efstathiou}}{1992}]{1992MNRAS.256P..43E}
{Efstathiou} G.,  1992, \mn@doi [\mnras] {10.1093/mnras/256.1.43P}, \href
  {http://adsabs.harvard.edu/abs/1992MNRAS.256P..43E} {256, 43P}

\bibitem[\protect\citeauthoryear{{Efstathiou}}{{Efstathiou}}{2000}]{2000MNRAS.317..697E}
{Efstathiou} G.,  2000, \mn@doi [\mnras] {10.1046/j.1365-8711.2000.03665.x},
  \href {http://adsabs.harvard.edu/abs/2000MNRAS.317..697E} {317, 697}

\bibitem[\protect\citeauthoryear{{Eisenstein} \& {Hut}}{{Eisenstein} \&
  {Hut}}{1998}]{HOP1998}
{Eisenstein} D.~J.,  {Hut} P.,  1998, \mn@doi [\apj] {10.1086/305535}, \href
  {http://adsabs.harvard.edu/abs/1998ApJ...498..137E} {498, 137}

\bibitem[\protect\citeauthoryear{{Escala} et~al.,}{{Escala}
  et~al.}{2018}]{Escala2018}
{Escala} I.,  et~al., 2018, \mn@doi [\mnras] {10.1093/mnras/stx2858}, \href
  {http://adsabs.harvard.edu/abs/2018MNRAS.474.2194E} {474, 2194}

\bibitem[\protect\citeauthoryear{{Fall} \& {Efstathiou}}{{Fall} \&
  {Efstathiou}}{1980}]{FallEfstathiou80}
{Fall} S.~M.,  {Efstathiou} G.,  1980, \mnras, \href
  {http://adsabs.harvard.edu/abs/1980MNRAS.193..189F} {193, 189}

\bibitem[\protect\citeauthoryear{{Faucher-Gigu{\`e}re}, {Lidz}, {Zaldarriaga}
  \& {Hernquist}}{{Faucher-Gigu{\`e}re} et~al.}{2009}]{FG2009}
{Faucher-Gigu{\`e}re} C.-A.,  {Lidz} A.,  {Zaldarriaga} M.,   {Hernquist} L.,
  2009, \mn@doi [\apj] {10.1088/0004-637X/703/2/1416}, \href
  {http://adsabs.harvard.edu/abs/2009ApJ...703.1416F} {703, 1416}

\bibitem[\protect\citeauthoryear{{Ferland}, {Korista}, {Verner}, {Ferguson},
  {Kingdon}  \& {Verner}}{{Ferland} et~al.}{1998}]{Ferland1998}
{Ferland} G.~J.,  {Korista} K.~T.,  {Verner} D.~A.,  {Ferguson} J.~W.,
  {Kingdon} J.~B.,   {Verner} E.~M.,  1998, \mn@doi [\pasp] {10.1086/316190},
  \href {http://adsabs.harvard.edu/abs/1998PASP..110..761F} {110, 761}

\bibitem[\protect\citeauthoryear{{Fisher} et~al.,}{{Fisher}
  et~al.}{2014}]{Fisher2014}
{Fisher} D.~B.,  et~al., 2014, \mn@doi [\nat] {10.1038/nature12765}, \href
  {http://adsabs.harvard.edu/abs/2014Natur.505..186F} {505, 186}

\bibitem[\protect\citeauthoryear{{Fitts} et~al.,}{{Fitts}
  et~al.}{2017}]{Fitts2017}
{Fitts} A.,  et~al., 2017, \mn@doi [\mnras] {10.1093/mnras/stx1757}, \href
  {http://adsabs.harvard.edu/abs/2017MNRAS.471.3547F} {471, 3547}

\bibitem[\protect\citeauthoryear{{Forbes}, {Krumholz}, {Goldbaum}  \&
  {Dekel}}{{Forbes} et~al.}{2016}]{Forbes2016}
{Forbes} J.~C.,  {Krumholz} M.~R.,  {Goldbaum} N.~J.,   {Dekel} A.,  2016,
  \mn@doi [\nat] {10.1038/nature18292}, \href
  {http://adsabs.harvard.edu/abs/2016Natur.535..523F} {535, 523}

\bibitem[\protect\citeauthoryear{{Garrison-Kimmel}, {Bullock}, {Boylan-Kolchin}
   \& {Bardwell}}{{Garrison-Kimmel} et~al.}{2017}]{Garrison-Kimmel2017}
{Garrison-Kimmel} S.,  {Bullock} J.~S.,  {Boylan-Kolchin} M.,   {Bardwell} E.,
  2017, \mn@doi [\mnras] {10.1093/mnras/stw2564}, \href
  {http://adsabs.harvard.edu/abs/2017MNRAS.464.3108G} {464, 3108}

\bibitem[\protect\citeauthoryear{{Gatto}, {Fraternali}, {Read}, {Marinacci},
  {Lux}  \& {Walch}}{{Gatto} et~al.}{2013}]{2013MNRAS.433.2749G}
{Gatto} A.,  {Fraternali} F.,  {Read} J.~I.,  {Marinacci} F.,  {Lux} H.,
  {Walch} S.,  2013, \mn@doi [\mnras] {10.1093/mnras/stt896}, \href
  {http://adsabs.harvard.edu/abs/2013MNRAS.433.2749G} {433, 2749}

\bibitem[\protect\citeauthoryear{{Geha} et~al.,}{{Geha}
  et~al.}{2013}]{Geha2013}
{Geha} M.,  et~al., 2013, \mn@doi [\apj] {10.1088/0004-637X/771/1/29}, \href
  {http://adsabs.harvard.edu/abs/2013ApJ...771...29G} {771, 29}

\bibitem[\protect\citeauthoryear{{Glover} \& {Clark}}{{Glover} \&
  {Clark}}{2012}]{Glover2012}
{Glover} S. C.~O.,  {Clark} P.~C.,  2012, \mn@doi [\mnras]
  {10.1111/j.1365-2966.2011.19648.x}, \href
  {https://ui.adsabs.harvard.edu/abs/2012MNRAS.421....9G} {421, 9}

\bibitem[\protect\citeauthoryear{{Gnedin} \& {Kaurov}}{{Gnedin} \&
  {Kaurov}}{2014}]{2014ApJ...793...30G}
{Gnedin} N.~Y.,  {Kaurov} A.~A.,  2014, \mn@doi [\apj]
  {10.1088/0004-637X/793/1/30}, \href
  {http://adsabs.harvard.edu/abs/2014ApJ...793...30G} {793, 30}

\bibitem[\protect\citeauthoryear{{Gnedin} \& {Kravtsov}}{{Gnedin} \&
  {Kravtsov}}{2006}]{2006ApJ...645.1054G}
{Gnedin} N.~Y.,  {Kravtsov} A.~V.,  2006, \mn@doi [\apj] {10.1086/504404},
  \href {http://adsabs.harvard.edu/abs/2006ApJ...645.1054G} {645, 1054}

\bibitem[\protect\citeauthoryear{{Gnedin} \& {Kravtsov}}{{Gnedin} \&
  {Kravtsov}}{2010}]{GnedinKravtsov2010}
{Gnedin} N.~Y.,  {Kravtsov} A.~V.,  2010, \mn@doi [\apj]
  {10.1088/0004-637X/714/1/287}, \href
  {http://adsabs.harvard.edu/abs/2010ApJ...714..287G} {714, 287}

\bibitem[\protect\citeauthoryear{{Gnedin}, {Tassis}  \& {Kravtsov}}{{Gnedin}
  et~al.}{2009}]{Gnedin09}
{Gnedin} N.~Y.,  {Tassis} K.,   {Kravtsov} A.~V.,  2009, \mn@doi [\apj]
  {10.1088/0004-637X/697/1/55}, \href
  {http://adsabs.harvard.edu/abs/2009ApJ...697...55G} {697, 55}

\bibitem[\protect\citeauthoryear{{Grebel}, {Gallagher}  \& {Harbeck}}{{Grebel}
  et~al.}{2003}]{2003AJ....125.1926G}
{Grebel} E.~K.,  {Gallagher} III J.~S.,   {Harbeck} D.,  2003, \mn@doi [\aj]
  {10.1086/368363}, \href {http://adsabs.harvard.edu/abs/2003AJ....125.1926G}
  {125, 1926}

\bibitem[\protect\citeauthoryear{{Grisdale}, {Agertz}, {Romeo}, {Renaud}  \&
  {Read}}{{Grisdale} et~al.}{2017}]{Grisdale2017}
{Grisdale} K.,  {Agertz} O.,  {Romeo} A.~B.,  {Renaud} F.,   {Read} J.~I.,
  2017, \mn@doi [\mnras] {10.1093/mnras/stw3133}, \href
  {http://adsabs.harvard.edu/abs/2017MNRAS.466.1093G} {466, 1093}

\bibitem[\protect\citeauthoryear{{Grisdale}, {Agertz}, {Renaud}  \&
  {Romeo}}{{Grisdale} et~al.}{2018}]{Grisdale2018}
{Grisdale} K.,  {Agertz} O.,  {Renaud} F.,   {Romeo} A.~B.,  2018, \mn@doi
  [\mnras] {10.1093/mnras/sty1595}, \href
  {http://adsabs.harvard.edu/abs/2018MNRAS.tmp.1523G} {}

\bibitem[\protect\citeauthoryear{{Grisdale}, {Agertz}, {Renaud}, {Romeo},
  {Devriendt}  \& {Slyz}}{{Grisdale} et~al.}{2019}]{Grisdale2019}
{Grisdale} K.,  {Agertz} O.,  {Renaud} F.,  {Romeo} A.~B.,  {Devriendt} J.,
  {Slyz} A.,  2019, arXiv e-prints, \href
  {http://adsabs.harvard.edu/abs/2019arXiv190200518G} {}

\bibitem[\protect\citeauthoryear{{Guillet} \& {Teyssier}}{{Guillet} \&
  {Teyssier}}{2011}]{GuilletTeyssier2011}
{Guillet} T.,  {Teyssier} R.,  2011, \mn@doi [Journal of Computational Physics]
  {10.1016/j.jcp.2011.02.044}, \href
  {http://adsabs.harvard.edu/abs/2011JCoPh.230.4756G} {230, 4756}

\bibitem[\protect\citeauthoryear{{Haardt} \& {Madau}}{{Haardt} \&
  {Madau}}{1996}]{haardtmadau96}
{Haardt} F.,  {Madau} P.,  1996, \mn@doi [\apj] {10.1086/177035}, \href
  {http://adsabs.harvard.edu/abs/1996ApJ...461...20H} {461, 20}

\bibitem[\protect\citeauthoryear{{Hanasz}, {Lesch}, {Naab}, {Gawryszczak},
  {Kowalik}  \& {W{\'o}lta{\'n}ski}}{{Hanasz} et~al.}{2013}]{Hanasz2013}
{Hanasz} M.,  {Lesch} H.,  {Naab} T.,  {Gawryszczak} A.,  {Kowalik} K.,
  {W{\'o}lta{\'n}ski} D.,  2013, \mn@doi [\apjl] {10.1088/2041-8205/777/2/L38},
  \href {http://adsabs.harvard.edu/abs/2013ApJ...777L..38H} {777, L38}

\bibitem[\protect\citeauthoryear{{Hopkins}, {Kere{\v s}}, {O{\~n}orbe},
  {Faucher-Gigu{\`e}re}, {Quataert}, {Murray}  \& {Bullock}}{{Hopkins}
  et~al.}{2014}]{Hopkins2014}
{Hopkins} P.~F.,  {Kere{\v s}} D.,  {O{\~n}orbe} J.,  {Faucher-Gigu{\`e}re}
  C.-A.,  {Quataert} E.,  {Murray} N.,   {Bullock} J.~S.,  2014, \mn@doi
  [\mnras] {10.1093/mnras/stu1738}, \href
  {http://adsabs.harvard.edu/abs/2014MNRAS.445..581H} {445, 581}

\bibitem[\protect\citeauthoryear{{Hopkins} et~al.,}{{Hopkins}
  et~al.}{2018}]{Hopkins2018}
{Hopkins} P.~F.,  et~al., 2018, \mn@doi [\mnras] {10.1093/mnras/sty1690}, \href
  {http://adsabs.harvard.edu/abs/2018MNRAS.480..800H} {480, 800}

\bibitem[\protect\citeauthoryear{{Hu}, {Naab}, {Glover}, {Walch}  \&
  {Clark}}{{Hu} et~al.}{2017}]{Hu2017}
{Hu} C.-Y.,  {Naab} T.,  {Glover} S.~C.~O.,  {Walch} S.,   {Clark} P.~C.,
  2017, \mn@doi [\mnras] {10.1093/mnras/stx1773}, \href
  {http://adsabs.harvard.edu/abs/2017MNRAS.471.2151H} {471, 2151}

\bibitem[\protect\citeauthoryear{{Iwamoto}, {Umeda}, {Tominaga}, {Nomoto}  \&
  {Maeda}}{{Iwamoto} et~al.}{2005}]{Iwamoto2005}
{Iwamoto} N.,  {Umeda} H.,  {Tominaga} N.,  {Nomoto} K.,   {Maeda} K.,  2005,
  \mn@doi [Science] {10.1126/science.1112997}, \href
  {http://adsabs.harvard.edu/abs/2005Sci...309..451I} {309, 451}

\bibitem[\protect\citeauthoryear{{Jaacks}, {Thompson}, {Finkelstein}  \&
  {Bromm}}{{Jaacks} et~al.}{2018}]{Jaacks2018}
{Jaacks} J.,  {Thompson} R.,  {Finkelstein} S.~L.,   {Bromm} V.,  2018, \mn@doi
  [\mnras] {10.1093/mnras/sty062}, \href
  {https://ui.adsabs.harvard.edu/\#abs/2018MNRAS.475.4396J} {475, 4396}

\bibitem[\protect\citeauthoryear{{Jeon}, {Bromm}, {Pawlik}  \&
  {Milosavljevi{\'c}}}{{Jeon} et~al.}{2015}]{Jeon2015}
{Jeon} M.,  {Bromm} V.,  {Pawlik} A.~H.,   {Milosavljevi{\'c}} M.,  2015,
  \mn@doi [\mnras] {10.1093/mnras/stv1353}, \href
  {http://adsabs.harvard.edu/abs/2015MNRAS.452.1152J} {452, 1152}

\bibitem[\protect\citeauthoryear{{Jethwa}, {Erkal}  \& {Belokurov}}{{Jethwa}
  et~al.}{2018}]{2018MNRAS.473.2060J}
{Jethwa} P.,  {Erkal} D.,   {Belokurov} V.,  2018, \mn@doi [\mnras]
  {10.1093/mnras/stx2330}, \href
  {https://ui.adsabs.harvard.edu/\#abs/2018MNRAS.473.2060J} {473, 2060}

\bibitem[\protect\citeauthoryear{{Katz}, {Weinberg}  \& {Hernquist}}{{Katz}
  et~al.}{1996}]{Katz1996}
{Katz} N.,  {Weinberg} D.~H.,   {Hernquist} L.,  1996, \mn@doi [\apjs]
  {10.1086/192305}, \href {http://adsabs.harvard.edu/abs/1996ApJS..105...19K}
  {105, 19}

\bibitem[\protect\citeauthoryear{{Kim} \& {Ostriker}}{{Kim} \&
  {Ostriker}}{2015}]{KimOstriker2015}
{Kim} C.-G.,  {Ostriker} E.~C.,  2015, \mn@doi [\apj]
  {10.1088/0004-637X/802/2/99}, \href
  {http://adsabs.harvard.edu/abs/2015ApJ...802...99K} {802, 99}

\bibitem[\protect\citeauthoryear{{Kimm}, {Katz}, {Haehnelt}, {Rosdahl},
  {Devriendt}  \& {Slyz}}{{Kimm} et~al.}{2017}]{Kimm2017}
{Kimm} T.,  {Katz} H.,  {Haehnelt} M.,  {Rosdahl} J.,  {Devriendt} J.,   {Slyz}
  A.,  2017, \mn@doi [\mnras] {10.1093/mnras/stx052}, \href
  {http://adsabs.harvard.edu/abs/2017MNRAS.466.4826K} {466, 4826}

\bibitem[\protect\citeauthoryear{{Kimm}, {Haehnelt}, {Blaizot}, {Katz},
  {Michel-Dansac}, {Garel}, {Rosdahl}  \& {Teyssier}}{{Kimm}
  et~al.}{2018}]{Kimm2018}
{Kimm} T.,  {Haehnelt} M.,  {Blaizot} J.,  {Katz} H.,  {Michel-Dansac} L.,
  {Garel} T.,  {Rosdahl} J.,   {Teyssier} R.,  2018, \mn@doi [\mnras]
  {10.1093/mnras/sty126}, \href
  {http://adsabs.harvard.edu/abs/2018MNRAS.475.4617K} {475, 4617}

\bibitem[\protect\citeauthoryear{{Kirby}, {Cohen}, {Guhathakurta}, {Cheng},
  {Bullock}  \& {Gallazzi}}{{Kirby} et~al.}{2013}]{Kirby2013}
{Kirby} E.~N.,  {Cohen} J.~G.,  {Guhathakurta} P.,  {Cheng} L.,  {Bullock}
  J.~S.,   {Gallazzi} A.,  2013, \mn@doi [\apj] {10.1088/0004-637X/779/2/102},
  \href {http://adsabs.harvard.edu/abs/2013ApJ...779..102K} {779, 102}

\bibitem[\protect\citeauthoryear{{Kirby}, {Bullock}, {Boylan-Kolchin},
  {Kaplinghat}  \& {Cohen}}{{Kirby} et~al.}{2014}]{Kirby2014}
{Kirby} E.~N.,  {Bullock} J.~S.,  {Boylan-Kolchin} M.,  {Kaplinghat} M.,
  {Cohen} J.~G.,  2014, \mn@doi [\mnras] {10.1093/mnras/stu025}, \href
  {http://adsabs.harvard.edu/abs/2014MNRAS.439.1015K} {439, 1015}

\bibitem[\protect\citeauthoryear{{Komatsu} et~al.,}{{Komatsu}
  et~al.}{2011}]{komatsu_etal11}
{Komatsu} E.,  et~al., 2011, \mn@doi [\apjs] {10.1088/0067-0049/192/2/18},
  \href {http://adsabs.harvard.edu/abs/2011ApJS..192...18K} {192, 18}

\bibitem[\protect\citeauthoryear{{Kravtsov}}{{Kravtsov}}{2003}]{2003ApJ...590L...1K}
{Kravtsov} A.~V.,  2003, \mn@doi [\apjl] {10.1086/376674}, \href
  {http://adsabs.harvard.edu/abs/2003ApJ...590L...1K} {590, L1}

\bibitem[\protect\citeauthoryear{{Kravtsov}, {Gnedin}  \& {Klypin}}{{Kravtsov}
  et~al.}{2004}]{2004ApJ...609..482K}
{Kravtsov} A.~V.,  {Gnedin} O.~Y.,   {Klypin} A.~A.,  2004, \mn@doi [\apj]
  {10.1086/421322}, \href {http://adsabs.harvard.edu/abs/2004ApJ...609..482K}
  {609, 482}

\bibitem[\protect\citeauthoryear{{Kroupa}}{{Kroupa}}{2001}]{Kroupa01}
{Kroupa} P.,  2001, \mn@doi [\mnras] {10.1046/j.1365-8711.2001.04022.x}, \href
  {http://adsabs.harvard.edu/abs/2001MNRAS.322..231K} {322, 231}

\bibitem[\protect\citeauthoryear{{Krumholz}}{{Krumholz}}{2012}]{Krumholz2012}
{Krumholz} M.~R.,  2012, \mn@doi [\apj] {10.1088/0004-637X/759/1/9}, \href
  {https://ui.adsabs.harvard.edu/abs/2012ApJ...759....9K} {759, 9}

\bibitem[\protect\citeauthoryear{{Krumholz} \& {Gnedin}}{{Krumholz} \&
  {Gnedin}}{2011}]{krumholzgnedin2011}
{Krumholz} M.~R.,  {Gnedin} N.~Y.,  2011, \mn@doi [\apj]
  {10.1088/0004-637X/729/1/36}, \href
  {http://adsabs.harvard.edu/abs/2011ApJ...729...36K} {729, 36}

\bibitem[\protect\citeauthoryear{{Krumholz} \& {Tan}}{{Krumholz} \&
  {Tan}}{2007}]{krumholztan07}
{Krumholz} M.~R.,  {Tan} J.~C.,  2007, \mn@doi [\apj] {10.1086/509101}, \href
  {http://adsabs.harvard.edu/abs/2007ApJ...654..304K} {654, 304}

\bibitem[\protect\citeauthoryear{{Lee}, {Miville-Desch{\^e}nes}  \&
  {Murray}}{{Lee} et~al.}{2016}]{Lee2016}
{Lee} E.~J.,  {Miville-Desch{\^e}nes} M.-A.,   {Murray} N.~W.,  2016, \mn@doi
  [\apj] {10.3847/1538-4357/833/2/229}, \href
  {https://ui.adsabs.harvard.edu/\#abs/2016ApJ...833..229L} {833, 229}

\bibitem[\protect\citeauthoryear{{Leitherer} et~al.,}{{Leitherer}
  et~al.}{1999}]{Leitherer1999}
{Leitherer} C.,  et~al., 1999, \mn@doi [\apjs] {10.1086/313233}, \href
  {http://adsabs.harvard.edu/abs/1999ApJS..123....3L} {123, 3}

\bibitem[\protect\citeauthoryear{{Leitner}}{{Leitner}}{2012}]{Leitner2012}
{Leitner} S.~N.,  2012, \mn@doi [\apj] {10.1088/0004-637X/745/2/149}, \href
  {http://adsabs.harvard.edu/abs/2012ApJ...745..149L} {745, 149}

\bibitem[\protect\citeauthoryear{{Levermore}}{{Levermore}}{1984}]{Levermore1984}
{Levermore} C.~D.,  1984, \mn@doi [\jqsrt] {10.1016/0022-4073(84)90112-2},
  \href {http://adsabs.harvard.edu/abs/1984JQSRT..31..149L} {31, 149}

\bibitem[\protect\citeauthoryear{{Macci{\`o}}, {Frings}, {Buck}, {Penzo},
  {Dutton}, {Blank}  \& {Obreja}}{{Macci{\`o}} et~al.}{2017}]{Maccio2017}
{Macci{\`o}} A.~V.,  {Frings} J.,  {Buck} T.,  {Penzo} C.,  {Dutton} A.~A.,
  {Blank} M.,   {Obreja} A.,  2017, \mn@doi [\mnras] {10.1093/mnras/stx2048},
  \href {http://adsabs.harvard.edu/abs/2017MNRAS.472.2356M} {472, 2356}

\bibitem[\protect\citeauthoryear{{Martizzi}, {Faucher-Gigu{\`e}re}  \&
  {Quataert}}{{Martizzi} et~al.}{2015}]{Martizzi2015}
{Martizzi} D.,  {Faucher-Gigu{\`e}re} C.-A.,   {Quataert} E.,  2015, \mn@doi
  [\mnras] {10.1093/mnras/stv562}, \href
  {http://adsabs.harvard.edu/abs/2015MNRAS.450..504M} {450, 504}

\bibitem[\protect\citeauthoryear{{Mashchenko}, {Wadsley}  \&
  {Couchman}}{{Mashchenko} et~al.}{2008}]{2008Sci...319..174M}
{Mashchenko} S.,  {Wadsley} J.,   {Couchman} H.~M.~P.,  2008, \mn@doi [Science]
  {10.1126/science.1148666}, \href
  {http://adsabs.harvard.edu/abs/2008Sci...319..174M} {319, 174}

\bibitem[\protect\citeauthoryear{{Mayer}, {Mastropietro}, {Wadsley}, {Stadel}
  \& {Moore}}{{Mayer} et~al.}{2006}]{2006MNRAS.369.1021M}
{Mayer} L.,  {Mastropietro} C.,  {Wadsley} J.,  {Stadel} J.,   {Moore} B.,
  2006, \mn@doi [\mnras] {10.1111/j.1365-2966.2006.10403.x}, \href
  {http://adsabs.harvard.edu/abs/2006MNRAS.369.1021M} {369, 1021}

\bibitem[\protect\citeauthoryear{{McConnachie}}{{McConnachie}}{2012}]{McConnachie2012}
{McConnachie} A.~W.,  2012, \mn@doi [\aj] {10.1088/0004-6256/144/1/4}, \href
  {http://adsabs.harvard.edu/abs/2012AJ....144....4M} {144, 4}

\bibitem[\protect\citeauthoryear{{Munshi} et~al.,}{{Munshi}
  et~al.}{2013}]{Munshi2013}
{Munshi} F.,  et~al., 2013, \mn@doi [\apj] {10.1088/0004-637X/766/1/56}, \href
  {http://adsabs.harvard.edu/abs/2013ApJ...766...56M} {766, 56}

\bibitem[\protect\citeauthoryear{{Munshi}, {Brooks}, {Applebaum}, {Weisz},
  {Governato}  \& {Quinn}}{{Munshi} et~al.}{2017}]{Munshi2017}
{Munshi} F.,  {Brooks} A.~M.,  {Applebaum} E.,  {Weisz} D.~R.,  {Governato} F.,
    {Quinn} T.~R.,  2017, preprint, \href
  {http://adsabs.harvard.edu/abs/2017arXiv170506286M} {} (\mn@eprint {arXiv}
  {1705.06286})

\bibitem[\protect\citeauthoryear{{Munshi}, {Brooks}, {Christensen},
  {Applebaum}, {Holley-Bockelmann}, {Quinn}  \& {Wadsley}}{{Munshi}
  et~al.}{2018}]{Munshi2018}
{Munshi} F.,  {Brooks} A.~M.,  {Christensen} C.,  {Applebaum} E.,
  {Holley-Bockelmann} K.,  {Quinn} T.~R.,   {Wadsley} J.,  2018, preprint,
  \href {http://adsabs.harvard.edu/abs/2018arXiv181012417M} {} (\mn@eprint
  {arXiv} {1810.12417})

\bibitem[\protect\citeauthoryear{{Murray}}{{Murray}}{2011}]{Murray2011b}
{Murray} N.,  2011, \mn@doi [\apj] {10.1088/0004-637X/729/2/133}, \href
  {http://adsabs.harvard.edu/abs/2011ApJ...729..133M} {729, 133}

\bibitem[\protect\citeauthoryear{{Nickerson}, {Teyssier}  \&
  {Rosdahl}}{{Nickerson} et~al.}{2018}]{Nickerson2018}
{Nickerson} S.,  {Teyssier} R.,   {Rosdahl} J.,  2018, \mn@doi [\mnras]
  {10.1093/mnras/sty1556}, \href
  {http://adsabs.harvard.edu/abs/2018MNRAS.tmp.1512N} {}

\bibitem[\protect\citeauthoryear{{O{\~n}orbe}, {Boylan-Kolchin}, {Bullock},
  {Hopkins}, {Kere{\v s}}, {Faucher-Gigu{\`e}re}, {Quataert}  \&
  {Murray}}{{O{\~n}orbe} et~al.}{2015}]{Onorbe2015}
{O{\~n}orbe} J.,  {Boylan-Kolchin} M.,  {Bullock} J.~S.,  {Hopkins} P.~F.,
  {Kere{\v s}} D.,  {Faucher-Gigu{\`e}re} C.-A.,  {Quataert} E.,   {Murray} N.,
   2015, \mn@doi [\mnras] {10.1093/mnras/stv2072}, \href
  {http://adsabs.harvard.edu/abs/2015MNRAS.454.2092O} {454, 2092}

\bibitem[\protect\citeauthoryear{{Ocvirk} et~al.,}{{Ocvirk}
  et~al.}{2018}]{2018arXiv181111192O}
{Ocvirk} P.,  et~al., 2018, arXiv e-prints, \href
  {http://adsabs.harvard.edu/abs/2018arXiv181111192O} {}

\bibitem[\protect\citeauthoryear{{Planck Collaboration} et~al.,}{{Planck
  Collaboration} et~al.}{2014}]{planck2014}
{Planck Collaboration} et~al., 2014, \mn@doi [\aap]
  {10.1051/0004-6361/201321591}, \href
  {http://adsabs.harvard.edu/abs/2014A%26A...571A..16P} {571, A16}

\bibitem[\protect\citeauthoryear{{Pontzen} \& {Governato}}{{Pontzen} \&
  {Governato}}{2012}]{2012MNRAS.421.3464P}
{Pontzen} A.,  {Governato} F.,  2012, \mn@doi [\mnras]
  {10.1111/j.1365-2966.2012.20571.x}, \href
  {http://adsabs.harvard.edu/abs/2012MNRAS.421.3464P} {421, 3464}

\bibitem[\protect\citeauthoryear{{Read} \& {Erkal}}{{Read} \&
  {Erkal}}{2018}]{Read2018}
{Read} J.~I.,  {Erkal} D.,  2018, preprint, \href
  {http://adsabs.harvard.edu/abs/2018arXiv180707093R} {} (\mn@eprint {arXiv}
  {1807.07093})

\bibitem[\protect\citeauthoryear{{Read} \& {Gilmore}}{{Read} \&
  {Gilmore}}{2005}]{2005MNRAS.356..107R}
{Read} J.~I.,  {Gilmore} G.,  2005, \mn@doi [\mnras]
  {10.1111/j.1365-2966.2004.08424.x}, \href
  {http://adsabs.harvard.edu/abs/2005MNRAS.356..107R} {356, 107}

\bibitem[\protect\citeauthoryear{{Read}, {Wilkinson}, {Evans}, {Gilmore}  \&
  {Kleyna}}{{Read} et~al.}{2006a}]{2006MNRAS.367..387R}
{Read} J.~I.,  {Wilkinson} M.~I.,  {Evans} N.~W.,  {Gilmore} G.,   {Kleyna}
  J.~T.,  2006a, \mn@doi [\mnras] {10.1111/j.1365-2966.2005.09959.x}, \href
  {http://adsabs.harvard.edu/abs/2006MNRAS.367..387R} {367, 387}

\bibitem[\protect\citeauthoryear{{Read}, {Pontzen}  \& {Viel}}{{Read}
  et~al.}{2006b}]{2006MNRAS.371..885R}
{Read} J.~I.,  {Pontzen} A.~P.,   {Viel} M.,  2006b, \mn@doi [\mnras]
  {10.1111/j.1365-2966.2006.10720.x}, \href
  {http://adsabs.harvard.edu/abs/2006MNRAS.371..885R} {371, 885}

\bibitem[\protect\citeauthoryear{{Read}, {Agertz}  \& {Collins}}{{Read}
  et~al.}{2016a}]{Read2016cores}
{Read} J.~I.,  {Agertz} O.,   {Collins} M.~L.~M.,  2016a, \mn@doi [\mnras]
  {10.1093/mnras/stw713}, \href
  {http://adsabs.harvard.edu/abs/2016MNRAS.459.2573R} {459, 2573}

\bibitem[\protect\citeauthoryear{{Read}, {Iorio}, {Agertz}  \&
  {Fraternali}}{{Read} et~al.}{2016b}]{Read2016galrot}
{Read} J.~I.,  {Iorio} G.,  {Agertz} O.,   {Fraternali} F.,  2016b, \mn@doi
  [\mnras] {10.1093/mnras/stw1876}, \href
  {http://adsabs.harvard.edu/abs/2016MNRAS.462.3628R} {462, 3628}

\bibitem[\protect\citeauthoryear{{Read}, {Iorio}, {Agertz}  \&
  {Fraternali}}{{Read} et~al.}{2017}]{Read2017}
{Read} J.~I.,  {Iorio} G.,  {Agertz} O.,   {Fraternali} F.,  2017, \mn@doi
  [\mnras] {10.1093/mnras/stx147}, \href
  {http://adsabs.harvard.edu/abs/2017MNRAS.467.2019R} {467, 2019}

\bibitem[\protect\citeauthoryear{{Revaz} \& {Jablonka}}{{Revaz} \&
  {Jablonka}}{2018}]{Revaz2018}
{Revaz} Y.,  {Jablonka} P.,  2018, \mn@doi [\aap]
  {10.1051/0004-6361/201832669}, \href
  {https://ui.adsabs.harvard.edu/\#abs/2018A&A...616A..96R} {616, A96}

\bibitem[\protect\citeauthoryear{{Rey} \& {Pontzen}}{{Rey} \&
  {Pontzen}}{2018}]{Rey18}
{Rey} M.~P.,  {Pontzen} A.,  2018, \mn@doi [\mnras] {10.1093/mnras/stx2744},
  \href {http://ukads.nottingham.ac.uk/abs/2018MNRAS.474...45R} {474, 45}

\bibitem[\protect\citeauthoryear{{Rey}, {Pontzen}, {Agertz}, {Orkney}, {Read},
  {Saintonge}  \& {Pedersen}}{{Rey} et~al.}{2019}]{Rey2019}
{Rey} M.~P.,  {Pontzen} A.,  {Agertz} O.,  {Orkney} M. D.~A.,  {Read} J.~I.,
  {Saintonge} A.,   {Pedersen} C.,  2019, \mn@doi [\apjl]
  {10.3847/2041-8213/ab53dd}, \href
  {https://ui.adsabs.harvard.edu/abs/2019ApJ...886L...3R} {886, L3}

\bibitem[\protect\citeauthoryear{{Rosdahl} \& {Blaizot}}{{Rosdahl} \&
  {Blaizot}}{2012}]{Rosdahl2012}
{Rosdahl} J.,  {Blaizot} J.,  2012, \mn@doi [\mnras]
  {10.1111/j.1365-2966.2012.20883.x}, \href
  {http://adsabs.harvard.edu/abs/2012MNRAS.423..344R} {423, 344}

\bibitem[\protect\citeauthoryear{{Rosdahl} \& {Teyssier}}{{Rosdahl} \&
  {Teyssier}}{2015}]{Rosdahl2015}
{Rosdahl} J.,  {Teyssier} R.,  2015, \mn@doi [\mnras] {10.1093/mnras/stv567},
  \href {http://adsabs.harvard.edu/abs/2015MNRAS.449.4380R} {449, 4380}

\bibitem[\protect\citeauthoryear{{Rosdahl}, {Blaizot}, {Aubert}, {Stranex}  \&
  {Teyssier}}{{Rosdahl} et~al.}{2013}]{Rosdahl2013}
{Rosdahl} J.,  {Blaizot} J.,  {Aubert} D.,  {Stranex} T.,   {Teyssier} R.,
  2013, \mn@doi [\mnras] {10.1093/mnras/stt1722}, \href
  {http://adsabs.harvard.edu/abs/2013MNRAS.436.2188R} {436, 2188}

\bibitem[\protect\citeauthoryear{{Rosdahl}, {Schaye}, {Teyssier}  \&
  {Agertz}}{{Rosdahl} et~al.}{2015}]{Rosdahl2015b}
{Rosdahl} J.,  {Schaye} J.,  {Teyssier} R.,   {Agertz} O.,  2015, \mn@doi
  [\mnras] {10.1093/mnras/stv937}, \href
  {https://ui.adsabs.harvard.edu/\#abs/2015MNRAS.451...34R} {451, 34}

\bibitem[\protect\citeauthoryear{{Rosdahl} et~al.,}{{Rosdahl}
  et~al.}{2018}]{Rosdahl2018}
{Rosdahl} J.,  et~al., 2018, \mn@doi [\mnras] {10.1093/mnras/sty1655}, \href
  {http://adsabs.harvard.edu/abs/2018MNRAS.479..994R} {479, 994}

\bibitem[\protect\citeauthoryear{{Rosen} \& {Bregman}}{{Rosen} \&
  {Bregman}}{1995}]{rosenbregman95}
{Rosen} A.,  {Bregman} J.~N.,  1995, \mn@doi [\apj] {10.1086/175303}, \href
  {http://adsabs.harvard.edu/abs/1995ApJ...440..634R} {440, 634}

\bibitem[\protect\citeauthoryear{{Roth}, {Pontzen}  \& {Peiris}}{{Roth}
  et~al.}{2016}]{Roth16}
{Roth} N.,  {Pontzen} A.,   {Peiris} H.~V.,  2016, \mn@doi [\mnras]
  {10.1093/mnras/stv2375}, \href
  {http://ukads.nottingham.ac.uk/abs/2016MNRAS.455..974R} {455, 974}

\bibitem[\protect\citeauthoryear{{Saitoh}, {Daisaka}, {Kokubo}, {Makino},
  {Okamoto}, {Tomisaka}, {Wada}  \& {Yoshida}}{{Saitoh}
  et~al.}{2008}]{2008PASJ...60..667S}
{Saitoh} T.~R.,  {Daisaka} H.,  {Kokubo} E.,  {Makino} J.,  {Okamoto} T.,
  {Tomisaka} K.,  {Wada} K.,   {Yoshida} N.,  2008, \mn@doi [\pasj]
  {10.1093/pasj/60.4.667}, \href
  {http://adsabs.harvard.edu/abs/2008PASJ...60..667S} {60, 667}

\bibitem[\protect\citeauthoryear{{Salem} \& {Bryan}}{{Salem} \&
  {Bryan}}{2014}]{SalemBryan2014}
{Salem} M.,  {Bryan} G.~L.,  2014, \mn@doi [\mnras] {10.1093/mnras/stt2121},
  \href {http://adsabs.harvard.edu/abs/2014MNRAS.437.3312S} {437, 3312}

\bibitem[\protect\citeauthoryear{{Sanders}, {Evans}  \& {Dehnen}}{{Sanders}
  et~al.}{2018}]{Sanders2018}
{Sanders} J.~L.,  {Evans} N.~W.,   {Dehnen} W.,  2018, \mn@doi [\mnras]
  {10.1093/mnras/sty1278}, \href
  {http://adsabs.harvard.edu/abs/2018MNRAS.478.3879S} {478, 3879}

\bibitem[\protect\citeauthoryear{{Sawala} et~al.,}{{Sawala}
  et~al.}{2015}]{Sawala2015}
{Sawala} T.,  et~al., 2015, \mn@doi [\mnras] {10.1093/mnras/stu2753}, \href
  {http://adsabs.harvard.edu/abs/2015MNRAS.448.2941S} {448, 2941}

\bibitem[\protect\citeauthoryear{{Simon}}{{Simon}}{2019}]{Simon2019}
{Simon} J.~D.,  2019, arXiv e-prints, \href
  {https://ui.adsabs.harvard.edu/\#abs/2019arXiv190105465S} {p.
  arXiv:1901.05465}

\bibitem[\protect\citeauthoryear{{Smith}, {Sijacki}  \& {Shen}}{{Smith}
  et~al.}{2019}]{SmithSijacki2019}
{Smith} M.~C.,  {Sijacki} D.,   {Shen} S.,  2019, \mn@doi [\mnras]
  {10.1093/mnras/stz599}, \href
  {https://ui.adsabs.harvard.edu/\#abs/2019MNRAS.485.3317S} {485, 3317}

\bibitem[\protect\citeauthoryear{{Springel}, {Frenk}  \& {White}}{{Springel}
  et~al.}{2006}]{Springel2006}
{Springel} V.,  {Frenk} C.~S.,   {White} S.~D.~M.,  2006, \mn@doi [\nat]
  {10.1038/nature04805}, \href
  {http://adsabs.harvard.edu/abs/2006Natur.440.1137S} {440, 1137}

\bibitem[\protect\citeauthoryear{{Stinson}, {Seth}, {Katz}, {Wadsley},
  {Governato}  \& {Quinn}}{{Stinson} et~al.}{2006a}]{2006MNRAS.373.1074S}
{Stinson} G.,  {Seth} A.,  {Katz} N.,  {Wadsley} J.,  {Governato} F.,   {Quinn}
  T.,  2006a, \mn@doi [\mnras] {10.1111/j.1365-2966.2006.11097.x}, \href
  {http://adsabs.harvard.edu/abs/2006MNRAS.373.1074S} {373, 1074}

\bibitem[\protect\citeauthoryear{{Stinson}, {Seth}, {Katz}, {Wadsley},
  {Governato}  \& {Quinn}}{{Stinson} et~al.}{2006b}]{Stinson06}
{Stinson} G.,  {Seth} A.,  {Katz} N.,  {Wadsley} J.,  {Governato} F.,   {Quinn}
  T.,  2006b, \mn@doi [\mnras] {10.1111/j.1365-2966.2006.11097.x}, \href
  {http://adsabs.harvard.edu/abs/2006MNRAS.373.1074S} {373, 1074}

\bibitem[\protect\citeauthoryear{{Stinson}, {Brook}, {Macci{\`o}}, {Wadsley},
  {Quinn}  \& {Couchman}}{{Stinson} et~al.}{2013}]{Stinson2013}
{Stinson} G.~S.,  {Brook} C.,  {Macci{\`o}} A.~V.,  {Wadsley} J.,  {Quinn}
  T.~R.,   {Couchman} H.~M.~P.,  2013, \mn@doi [\mnras] {10.1093/mnras/sts028},
  \href {http://adsabs.harvard.edu/abs/2013MNRAS.428..129S} {428, 129}

\bibitem[\protect\citeauthoryear{{Teyssier}}{{Teyssier}}{2002}]{teyssier02}
{Teyssier} R.,  2002, \mn@doi [\aap] {10.1051/0004-6361:20011817}, \href
  {http://adsabs.harvard.edu/abs/2002A%26A...385..337T} {385, 337}

\bibitem[\protect\citeauthoryear{{Tollet} et~al.,}{{Tollet}
  et~al.}{2016}]{Tollet2016}
{Tollet} E.,  et~al., 2016, \mn@doi [\mnras] {10.1093/mnras/stv2856}, \href
  {http://adsabs.harvard.edu/abs/2016MNRAS.456.3542T} {456, 3542}

\bibitem[\protect\citeauthoryear{{Toro}, {Spruce}  \& {Speares}}{{Toro}
  et~al.}{1994}]{Toro1994}
{Toro} E.~F.,  {Spruce} M.,   {Speares} W.,  1994, \mn@doi [Shock Waves]
  {10.1007/BF01414629}, \href
  {http://adsabs.harvard.edu/abs/1994ShWav...4...25T} {4, 25}

\bibitem[\protect\citeauthoryear{{Torrey}, {Vogelsberger}, {Genel}, {Sijacki},
  {Springel}  \& {Hernquist}}{{Torrey} et~al.}{2014}]{2014MNRAS.438.1985T}
{Torrey} P.,  {Vogelsberger} M.,  {Genel} S.,  {Sijacki} D.,  {Springel} V.,
  {Hernquist} L.,  2014, \mn@doi [\mnras] {10.1093/mnras/stt2295}, \href
  {http://adsabs.harvard.edu/abs/2014MNRAS.438.1985T} {438, 1985}

\bibitem[\protect\citeauthoryear{{Vandenbroucke}, {Verbeke}  \& {De
  Rijcke}}{{Vandenbroucke} et~al.}{2016}]{Vandenbrouke2016}
{Vandenbroucke} B.,  {Verbeke} R.,   {De Rijcke} S.,  2016, \mn@doi [\mnras]
  {10.1093/mnras/stw328}, \href
  {http://adsabs.harvard.edu/abs/2016MNRAS.458..912V} {458, 912}

\bibitem[\protect\citeauthoryear{Verbeke, Vandenbroucke  \& Rijcke}{Verbeke
  et~al.}{2015}]{Verbeke2015}
Verbeke R.,  Vandenbroucke B.,   Rijcke S.~D.,  2015, \mn@doi [The
  Astrophysical Journal] {10.1088/0004-637x/815/2/85}, 815, 85

\bibitem[\protect\citeauthoryear{{Vikhlinin} et~al.,}{{Vikhlinin}
  et~al.}{2009}]{vikhlinin_etal09b}
{Vikhlinin} A.,  et~al., 2009, \mn@doi [\apj] {10.1088/0004-637X/692/2/1060},
  \href {http://adsabs.harvard.edu/abs/2009ApJ...692.1060V} {692, 1060}

\bibitem[\protect\citeauthoryear{{Weisz}, {Dolphin}, {Skillman}, {Holtzman},
  {Gilbert}, {Dalcanton}  \& {Williams}}{{Weisz}
  et~al.}{2014}]{2014ApJ...789..148W}
{Weisz} D.~R.,  {Dolphin} A.~E.,  {Skillman} E.~D.,  {Holtzman} J.,  {Gilbert}
  K.~M.,  {Dalcanton} J.~J.,   {Williams} B.~F.,  2014, \mn@doi [\apj]
  {10.1088/0004-637X/789/2/148}, \href
  {http://adsabs.harvard.edu/abs/2014ApJ...789..148W} {789, 148}

\bibitem[\protect\citeauthoryear{{Wetzel}, {Hopkins}, {Kim},
  {Faucher-Gigu{\`e}re}, {Kere{\v{s}}}  \& {Quataert}}{{Wetzel}
  et~al.}{2016}]{Wetzel2016}
{Wetzel} A.~R.,  {Hopkins} P.~F.,  {Kim} J.-h.,  {Faucher-Gigu{\`e}re} C.-A.,
  {Kere{\v{s}}} D.,   {Quataert} E.,  2016, \mn@doi [\apj]
  {10.3847/2041-8205/827/2/L23}, \href
  {https://ui.adsabs.harvard.edu/\#abs/2016ApJ...827L..23W} {827, L23}

\bibitem[\protect\citeauthoryear{{Wheeler}, {O{\~n}orbe}, {Bullock},
  {Boylan-Kolchin}, {Elbert}, {Garrison-Kimmel}, {Hopkins}  \& {Kere{\v
  s}}}{{Wheeler} et~al.}{2015}]{Wheeler2015}
{Wheeler} C.,  {O{\~n}orbe} J.,  {Bullock} J.~S.,  {Boylan-Kolchin} M.,
  {Elbert} O.~D.,  {Garrison-Kimmel} S.,  {Hopkins} P.~F.,   {Kere{\v s}} D.,
  2015, \mn@doi [\mnras] {10.1093/mnras/stv1691}, \href
  {http://adsabs.harvard.edu/abs/2015MNRAS.453.1305W} {453, 1305}

\bibitem[\protect\citeauthoryear{{Wheeler} et~al.,}{{Wheeler}
  et~al.}{2018}]{Wheeler2018}
{Wheeler} C.,  et~al., 2018, arXiv e-prints, \href
  {http://adsabs.harvard.edu/abs/2018arXiv181202749W} {}

\bibitem[\protect\citeauthoryear{{White} \& {Rees}}{{White} \&
  {Rees}}{1978}]{WhiteRees78}
{White} S.~D.~M.,  {Rees} M.~J.,  1978, \mnras, \href
  {http://adsabs.harvard.edu/abs/1978MNRAS.183..341W} {183, 341}

\bibitem[\protect\citeauthoryear{{Wiersma}, {Schaye}, {Theuns}, {Dalla Vecchia}
   \& {Tornatore}}{{Wiersma} et~al.}{2009}]{Wiersma2009}
{Wiersma} R.~P.~C.,  {Schaye} J.,  {Theuns} T.,  {Dalla Vecchia} C.,
  {Tornatore} L.,  2009, \mn@doi [\mnras] {10.1111/j.1365-2966.2009.15331.x},
  \href {http://adsabs.harvard.edu/abs/2009MNRAS.399..574W} {399, 574}

\bibitem[\protect\citeauthoryear{{Wise}, {Abel}, {Turk}, {Norman}  \&
  {Smith}}{{Wise} et~al.}{2012a}]{Wise2012}
{Wise} J.~H.,  {Abel} T.,  {Turk} M.~J.,  {Norman} M.~L.,   {Smith} B.~D.,
  2012a, \mn@doi [\mnras] {10.1111/j.1365-2966.2012.21809.x}, \href
  {http://adsabs.harvard.edu/abs/2012MNRAS.427..311W} {427, 311}

\bibitem[\protect\citeauthoryear{{Wise}, {Turk}, {Norman}  \& {Abel}}{{Wise}
  et~al.}{2012b}]{wise_etal12}
{Wise} J.~H.,  {Turk} M.~J.,  {Norman} M.~L.,   {Abel} T.,  2012b, \mn@doi
  [\apj] {10.1088/0004-637X/745/1/50}, \href
  {http://adsabs.harvard.edu/abs/2012ApJ...745...50W} {745, 50}

\bibitem[\protect\citeauthoryear{{Wise}, {Demchenko}, {Halicek}, {Norman},
  {Turk}, {Abel}  \& {Smith}}{{Wise} et~al.}{2014}]{Wise2014}
{Wise} J.~H.,  {Demchenko} V.~G.,  {Halicek} M.~T.,  {Norman} M.~L.,  {Turk}
  M.~J.,  {Abel} T.,   {Smith} B.~D.,  2014, \mn@doi [\mnras]
  {10.1093/mnras/stu979}, \href
  {http://adsabs.harvard.edu/abs/2014MNRAS.442.2560W} {442, 2560}

\bibitem[\protect\citeauthoryear{{Woosley} \& {Weaver}}{{Woosley} \&
  {Weaver}}{1995}]{woosleyweaver1995}
{Woosley} S.~E.,  {Weaver} T.~A.,  1995, \mn@doi [\apjs] {10.1086/192237},
  \href {http://adsabs.harvard.edu/abs/1995ApJS..101..181W} {101, 181}

\bibitem[\protect\citeauthoryear{{Wright}, {Brooks}, {Weisz}  \&
  {Christensen}}{{Wright} et~al.}{2018}]{Wright2018}
{Wright} A.~C.,  {Brooks} A.~M.,  {Weisz} D.~R.,   {Christensen} C.~R.,  2018,
  \mn@doi [\mnras] {10.1093/mnras/sty2759}, \href
  {http://adsabs.harvard.edu/abs/2018MNRAS.tmp.2641W} {}

\makeatother
\end{thebibliography}

}

\end{document}